\documentstyle[11pt,aaspp4,flushrt]{article}

\def\Lya{Ly$\alpha\ $}

\def\HI{\hbox{H~$\rm \scriptstyle I\ $}}
\def\HII{\hbox{H~$\rm \scriptstyle II\ $}}

\def\HeI{\hbox{He~$\rm \scriptstyle I\ $}}
\def\HeII{\hbox{He~$\rm \scriptstyle II\ $}}
\def\HeIII{\hbox{He~$\rm \scriptstyle III\ $}}

\def\NHI{N_{\rm HI}}
\def\NHeII{N_{\rm HeII}}
\def\msun{\,{\rm M_\odot}\,}
\def\cm2{\,{\rm cm^{-2}}\,}
\def\pcc{\,{\rm cm^{-3}}\,}
\def\kms{\,{\rm km\,s^{-1}}\,}
\def\kmsmpc{\,{\rm km\,s^{-1}\,Mpc^{-1}}\,}
\def\kpc{\,{\rm kpc\,}}
\def\eV{\,{\rm eV\ }}
\def\kel{\,{\rm K\ }}

\def\ltsima{$\; \buildrel < \over \sim \;$}
\def\lsim{\lower.5ex\hbox{\ltsima}}
\def\gtsima{$\; \buildrel > \over \sim \;$}
\def\gsim{\lower.5ex\hbox{\gtsima}}
\def\etal{{ et~al.~}}

\def\mur86{Murdoch \etal 1986}
\def\hu95{Hu \etal 1995}
\def\lu91{Lu \etal 1991}
\def\luu96{Lu \etal 1997}
\def\gia91{Giallongo 1991}
\def\bec94{Bechtold 1994}
\def\bah93{Bahcall \etal 1993}
\def\bahh96{Bahcall \etal 1996}

\begin{document}

\title{Physical Properties of the Lyman Alpha Forest
       in a Cold Dark Matter Cosmology}

\author{Yu Zhang$^{1,2}$,
            Avery Meiksin$^{3,4}$,
            Peter Anninos$^{1}$, and
            Michael L. Norman$^{1,2}$}

\vskip10pt
\affil{\em $^1$Laboratory for Computational Astrophysics, National Center
for Supercomputing Applications, University of Illinois at Urbana-Champaign,
405 N. Mathews Ave., Urbana, IL\ 61801}

\vskip10pt
\affil{\em $^2$Astronomy Department, University of Illinois at Urbana-
Champaign, 1002 West Green Street, Urbana, IL \ 61801}

\vskip10pt
\affil{\em $^3$Department of Astronomy \& Astrophysics, University of
Chicago, 5640 South Ellis Avenue, Chicago, IL\ 60637}

\vskip10pt
\affil{\em $^4$Edwin P. Hubble Research Scientist}

\date{\today}

\begin{abstract}

We discuss the origin and physical nature of the \Lya forest
absorption systems as found in hydrodynamical simulations of the
Intergalactic Medium (IGM) in a standard Cold Dark Matter cosmology
($\Omega=1$, $H_0=50\kmsmpc$, $\sigma_8=0.7$). The structures of the
systems that give rise to the \Lya forest span a wide range in
morphologies, depending on the density contrast. Highly overdense
systems, $\rho_b/\bar\rho_b\gsim10$, where $\rho_b$ denotes baryon
density, tend to be spheroidal and are located at the intersections of
an interconnecting network of filaments of moderate overdensity,
$1\lsim\rho_b/\bar\rho_b\lsim5$. The typical thickness of the
filaments is 100 kpc, with a typical length of a few megaparsecs. At
the cosmological average density, the characteristic morphology is
cell--like with underdense regions separated by overdense sheet--like
partitions. The lowest density contours tend to enclose amorphous,
isolated regions. We find that the principal structures of the IGM are
in place by $z=5$, with the evolution in the IGM absorption properties
due primarily to the expansion of the universe and the changing
intensity of the photoionizing background radiation field. The
absorption properties of the forest clouds correlate strongly with
those of the underlying physical systems from which they arise. The
highest column density systems ($\log\NHI\gsim15$), correspond to the
highly overdense spheroidal structures, moderate column density
systems ($13\lsim\log\NHI\lsim14$), correspond to the filaments, and
the lowest density absorption systems originate from discrete
fluctuations within underdense regions a few megaparsecs across,
cosmic minivoids. Most of the intergalactic \HeII opacity arises from
these underdense regions. Similar correlations are found for the
cloud temperature and divergence of the peculiar velocity field.
Within the uncertainties of the statistics of the derived \Lya forest
properties, we are able to account for the distribution of optical
depths in our synthesized spectra entirely by absorption due to
discrete systems. We find that virtually all the baryons in the
simulation fragment into structures that we can identify with discrete
absorption lines, with at most 5\% remaining in a smoothly distributed
component (the Gunn--Peterson effect). We compare our results with the
cloud ionization parameters inferred from Keck HIRES measurements of
carbon and silicon in the \Lya forest. Combining with constraints
imposed by measurements of the mean intergalactic \HI opacity permits
separate limits to be set on the mean cosmological baryon density
$\Omega_b$ and \HI ionization rate $\Gamma_{\rm HI}$. For the
cosmological model investigated, we find $0.03\lsim\Omega_b\lsim0.08$
and $0.3\lsim\Gamma_{\rm HI,-12}\lsim1$ ($\Gamma_{\rm
HI,-12}=\Gamma_{\rm HI}/ 10^{-12}\, {\rm s^{-1}}$), at $z=3-3.5$. Our
results for the amount of intergalactic \HI and \HeII absorption and
for the ionization parameters of the clouds are consistent with a forest
photoionized by a UV background dominated by QSO sources
with an intrinsic spectral index of $\alpha_Q\approx1.8-2$.

\end{abstract}


\newpage

\section{Introduction}
\label{intro}

The last few years have witnessed considerable advances in our
understanding of the structure of the Intergalactic Medium (IGM)
predicted by Cold Dark Matter (CDM) dominated cosmologies. Several
groups have performed a series of combined numerical $N$-body/
hydrodynamics simulations of structure formation in the IGM (Cen \etal
1994; Zhang, Anninos, \& Norman 1995; Hernquist
\etal 1996), that are converging on a definite picture for the origin
of the \Lya forest in a CDM universe.  Although some differences
between the simulations remain to be resolved, the general landscape
the simulations have drawn is one of an interconnected network of
sheets and filaments, with dwarfish spheroidal systems, essentially
minihalos (Rees 1986; Ikeuchi 1986), located at their points of
intersection, and fluctuations within low density regions between. The
filamentary structure bears a remarkable resemblance to the findings
of similar simulations for the formation of rich clusters of galaxies
(Bryan \etal 1994).  Bond, Kofman, \& Pogosyan (1996) have argued that
this ``cosmic web'' is a generic feature of CDM, a consequence of an
inchoate pattern in the matter fluctuations imprinted at the epoch of
matter--radiation decoupling and sharpened by gravitational instability.
The distribution in neutral hydrogen column densities along
lines--of--sight piercing the filaments, as well as the distribution
of the velocity widths of the resulting absorption features, are found
to coincide closely with the measurements of the \Lya forest in the
spectra of high redshift quasars (Miralda-Escud\'e \etal 1996; Dav\'e
\etal 1997; Zhang \etal 1997; Wadsley \& Bond 1997).

Ever since the pioneering survey of Sargent \etal (1980),
observational studies of the \Lya forest have targeted several key
issues concerning the structure of the absorbers:\ 1.\ What is the
physical state of the clouds?\ 2.\ What is the confinement mechanism
of the clouds?\ 3.\ How do the clouds evolve?\ 4.\ What are their
shapes and physical extent?\ 5.\ How do the clouds fit into scenarios
for the formation of large--scale structure? Following the
identification of the \Lya absorption lines in QSO spectra as
intergalactic \HI clouds by Lynds (1971), Arons (1972) attributed the
absorption to the hydrogen in intervening protogalaxies photoionized
by a QSO--dominated UV background. The essential confinement mechanism
in this scenario is the gravity of the protogalaxy. Sargent \etal
(1980) suggested an alternative model in which the absorption arises
from intergalactic gas clouds confined by a hot ambient IGM. Following
their lead, Ostriker \& Ikeuchi (1983) and Ikeuchi \& Ostriker (1986)
formulated a theory of pressure confined clouds that permitted
definite predictions to be made for the observed properties of the
absorbers and their evolution and for the properties of the hot confining
medium. Shortly thereafter, the successes of the CDM model created an
interest in combining this theory of large scale structure formation
with the formation of \Lya clouds into a single unified picture. In
this scenario, \Lya clouds can either be associated with bound
structures, stabilized by the gravity of dark matter mini--halos (Rees
1986, 1988; Ikeuchi 1986), or with post--photoionized unconfined gas
in low mass objects that developed from small mass fluctuations (Bond,
Szalay, and Silk 1988). Subsequent discussions of the structure and
evolution of the clouds included a link to dwarf galaxy formation
(Ikeuchi \& Norman 1987; Ikeuchi, Murakami, \& Rees 1988), the effect
of environment on minihalos (Murakami \& Ikeuchi 1994), and slab
models for the clouds (McGill 1990; Charlton, Salpeter, \& Hogan 1993;
Miralda--Escud\'e \& Rees 1993; Meiksin 1994).

Parallel to the investigations of the \Lya forest has been a closely
allied effort to study the structure of a smoothly distributed diffuse
intergalactic gas component. Soon after the identification of the
first QSO, Gunn \& Peterson (1965) recognized that its spectrum could
be used to place a stringent constraint on the number density of
neutral hydrogen in the IGM by measuring the amount of \Lya absorption
shortward of the \Lya emission line of the QSO. The absence of a
strong absorption trough led them to conclude that the neutral
hydrogen density of the IGM is so low that the IGM must be highly
ionized. The detection of such a component would have important
consequences for cosmology. Since a lower limit to the amount of
metagalactic ionizing radiation is provided by the contribution due to
QSO sources, a measurement of the \Lya opacity of a smooth component
could be used to determine a lower bound to the total hydrogen density
of the IGM, assuming the IGM is in photoionization equilibrium. This
density would be directly comparable to the prediction of Big Bang
nucleosynthesis. In order to estimate the opacity of the smooth
component, however, it is crucial first to remove the contribution to
the absorption due to the \Lya forest (Steidel \& Sargent 1987;
Jenkins \& Ostriker 1991). The total density of the diffuse gas may be
derived only if its filling factor is known, which, following Gunn \&
Peterson (1965), was implicitly assumed to
be unity. For a clumped component like the forest, the filling factor
is not directly amenable to measurements. To date there has been no
definitive detection of a smoothly distributed component of neutral
hydrogen. Measurements have been made of the amount of absorption by
intergalactic \HeII (Jakobsen \etal 1994; Davidsen \etal 1996;
Hogan \etal 1997), but again it appears possible to account for the
absorption entirely by the \Lya forest (Madau \& Meiksin 1994;
Songaila, Hu, \& Cowie 1995; Giroux, Fardal, \& Shull 1995).

In this paper, we address these questions using the results of our
recent numerical hydrodynamics simulations of the formation of the
\Lya forest in standard CDM (SCDM). In a companion paper, Zhang \etal
(1997), we presented an analysis of synthetic spectra constructed from
the simulation results. There we found excellent agreement between the
model \HI column density distribution over the range $10^{12}<\NHI<10^{16}\cm2$
and the observed spectra from the Keck HIRES and from 
previous high spectral resolution
studies. We also showed the underlying distribution of Doppler parameters
agrees closely with the distribution inferred from observations.
We found that agreement with measurements of the intergalactic \ion{He}{2}
absorption required a fairly soft UV background, but one still compatible
with QSOs as the dominant radiation sources.
In this paper, we concentrate on the physical structure of the clouds.
We make some comparisons with the results from alternative
schemes for solving the hydrodynamics. The calculations of
Hernquist \etal (1996), Haehnelt \etal (1996), and Wadsley \& Bond (1997)
were performed using
Smoothed Particle Hydrodynamics (SPH), in contrast to the more traditional
grid--based code we have adopted. 
Cen \etal (1994) performed a $\Lambda$CDM simulation
using a grid--based code, though with a different treatment of
the hydrodynamics based on a total variation dimininishing scheme (TVD).
We also provide a detailed discussion of a topic that has not received much
previous attention, the structure within small
underdense regions, minivoids. We shall argue that an understanding
of the fine--structure of the minivoids is essential for interpreting
measurements of the opacity of the IGM, both of hydrogen and especially
of singly ionized helium.

In \S \ref{sec:sim} we briefly describe our numerical method and
simulations. In \S \ref{sec:physical} we describe the physical properties
of the absorption systems, and relate these to their absorption properties.
We discuss the origin of the absorption in the underdense regions in
\S \ref{sec:minivoids} and its observational consequences.
We summarize our results in \S \ref{sec:summary}.

\section{The Simulations}
\label{sec:sim}

The simulations were performed using our 2--level hierarchical grid
code HERCULES (Anninos, Norman, \& Clarke 1994; Anninos \etal 1997).
The simulation box had a comoving side of 9.6 Mpc. Most of the
analysis presented here is for the top grid. To explore the effects of
grid resolution, we take advantage of the 2--level nature of our code
and introduce a second, more finely resolved, subgrid to cover a
section of the coarser top grid. We use $128^3$ cells on both the top
and subgrids, and $128^3$ particles to represent the dark matter. The
subgrid is centered on the {\it least} dense region of the top grid
for the purpose of resolving in greater detail the fine density
structure of the voids. The subgrid is a factor of 4 greater in
resolution. Thus, the top grid has a comoving resolution of 75 kpc and
the subgrid of 18.75 kpc. The dark matter particles have a mass of
$2.9\times10^7\msun$ in the top grid, and $4.6\times10^5\msun$ in the subgrid.

Our model background spacetime is a flat, cold dark matter dominated universe
with the initial density perturbations originating from inflation--inspired
adiabatic fluctuations. We assume a Hubble constant $H_0=50\kmsmpc$.
The BBKS (Bardeen \etal 1986) transfer function is employed with the standard
Harrison--Zel'dovich power spectrum, normalized to $\sigma_{8h^{-1}}=0.7$,
where $h=H_0/100\kmsmpc$, consistent with the present number
density and temperatures of galaxy clusters (White, Efstathiou, \& Frenk 1993;
Bond \& Myers 1996). We adopt $\Omega_b=0.06$, consistent with Big Bang
nucleosynthesis limits (Copi, Schramm, \& Turner 1995), and the
baryonic fluid is composed of hydrogen and helium in primordial abundance
with a hydrogen mass fraction of 76\%. We generate the initial particle
positions and perturbations using the COSmological initial conditions
and MICrowave anisotropy codeS (COSMICS, Bertschinger 1995).

In addition to the usual ingredients of baryonic and dark matter, we
also solve the coupled system of non--equilibrium chemical reactions
with radiative cooling. The reaction network includes a
self--consistent treatment of the following six species: \HI, \HII,
\HeI, \HeII, \HeIII and $e^-$. Details of the chemical model, cooling
rates and numerical methods can be found in Abel \etal (1997) and
Anninos \etal (1997). We include a uniform UV photoionizing
background, adopting the estimate of Haardt \& Madau (1996) for a
QSO-dominated UV background. We turn on the radiation field at $z=6$.
It is unlikely that the \HII regions from QSO sources would have percolated
before this time (Meiksin \& Madau 1993), though it is possible the IGM was
reionized earlier by other sources like an early generation
of stars or decaying neutrinos. The evolution of systems sufficiently dense
as to be in photoionization thermal equilibrium are not expected to be much
affected by the turn on time of the radiation field by $z<5$, but more
rarefied systems may be sensitive to the reionization history (Meiksin 1994).
The topic merits further investigation. We do not incorporate radiative
transfer through the gas, as the clouds we are most interested in have column
densities $\NHI<10^{16}\cm2$. Radiative transfer of the \HI ionizing
radiation becomes significant for column densities a factor of several
above this. For \HeII, radiative transfer is just starting to become
important for this column density, so that we may slightly
overestimate the temperatures of the higher column density systems,
but the dynamics should be essentially unaltered at our resolution.
We adopt the fits to the photoionization rates from Haardt \& Madau
(1996), and the fits to their results for the heating rates presented in
Zhang \etal (1997). Since the cloud temperatures are not very
sensitive to the assumed heating rate, we note that comparisons of our
current results with alternative radiation fields or cosmic baryon
densities may be made by simply rescaling the ionization fractions using
the ratio of parameters $b_{\rm ion}\equiv\Omega_b^2/\Gamma_{-12}$, where
$\Gamma_{-12}$ is the photoionization rate for the species of interest,
in units of $10^{-12}s^{-1}$. This is a valid procedure provided the
regions have not undergone significant recombination into neutral
hydrogen or neutral or singly ionized helium, in which case the accompanying
change in the temperatures of the clouds can have a non--negligible effect
on the cloud dynamics and structure.\footnote{Because of the
temperature dependence of the radiative recombination rates, a slight
dependence of the gas temperature on the baryon density reduces the scaling
of the ionization fraction to being somewhat weaker than in direct
proportion to ${\rm b_{ion}}$.}

\section{Physical Properties of Absorbers}
\label{sec:physical}

\subsection{Morphologies}
\label{subsec:morph}

The cloud population is characterized by a range of morphologies. The
highest column density systems tend to be associated with physically
more dense and compact structures, similar to the minihalo model. These
spheroidal systems are interconnected by a network of sheets and filaments.
It is these structures that are responsible for most of the traditional \Lya
forest, with column densities in the range $10^{13}<\NHI<10^{15}\cm2$.
In between the filaments are underdense regions. Even though underdense,
these regions give rise to discrete absorption systems as well. Because
the underlying physical origin of these systems appears distinct from
those above, and because of its relevance to the detection of a smooth diffuse
IGM component, we defer a more complete discussion of the systems in the
underdense regions until the following section.

Although no unique attribute suffices to describe the full range of
morphologies exhibited by the clouds, systems of a given baryon
overdensity are associated with distinctive shapes and configurations.
In Plate 1, we show the configurations enclosed by 3D isodensity contours
at $z=3$. At low densities (first panel, $\rho/\bar\rho=0.1$), the
contours enclose isolated amorphous regions. These lie at the centers
of minivoids, underdense regions a few (comoving) megaparsecs across,
as shown in the second and third panels ($\rho/\bar\rho=0.3$ and
$\rho/\bar\rho=0.5$). The systems corresponding to the cosmological average
density ($\rho/\bar\rho=1$), in the fourth panel show a cell--like or spongy
morphology, with underdense regions separated by wall--like partitions
of varying thicknesses. The next panel ($\rho/\bar\rho=3$), shows a filamentary
network emerging from the intersections of the walls. The network, or cosmic
web, is prominent in the final panel ($\rho/\bar\rho=10$). We have provided
an accompanying video that more vividly displays the transitions between
the various morphologies and their interrelations.

\begin{figure}
\plotone{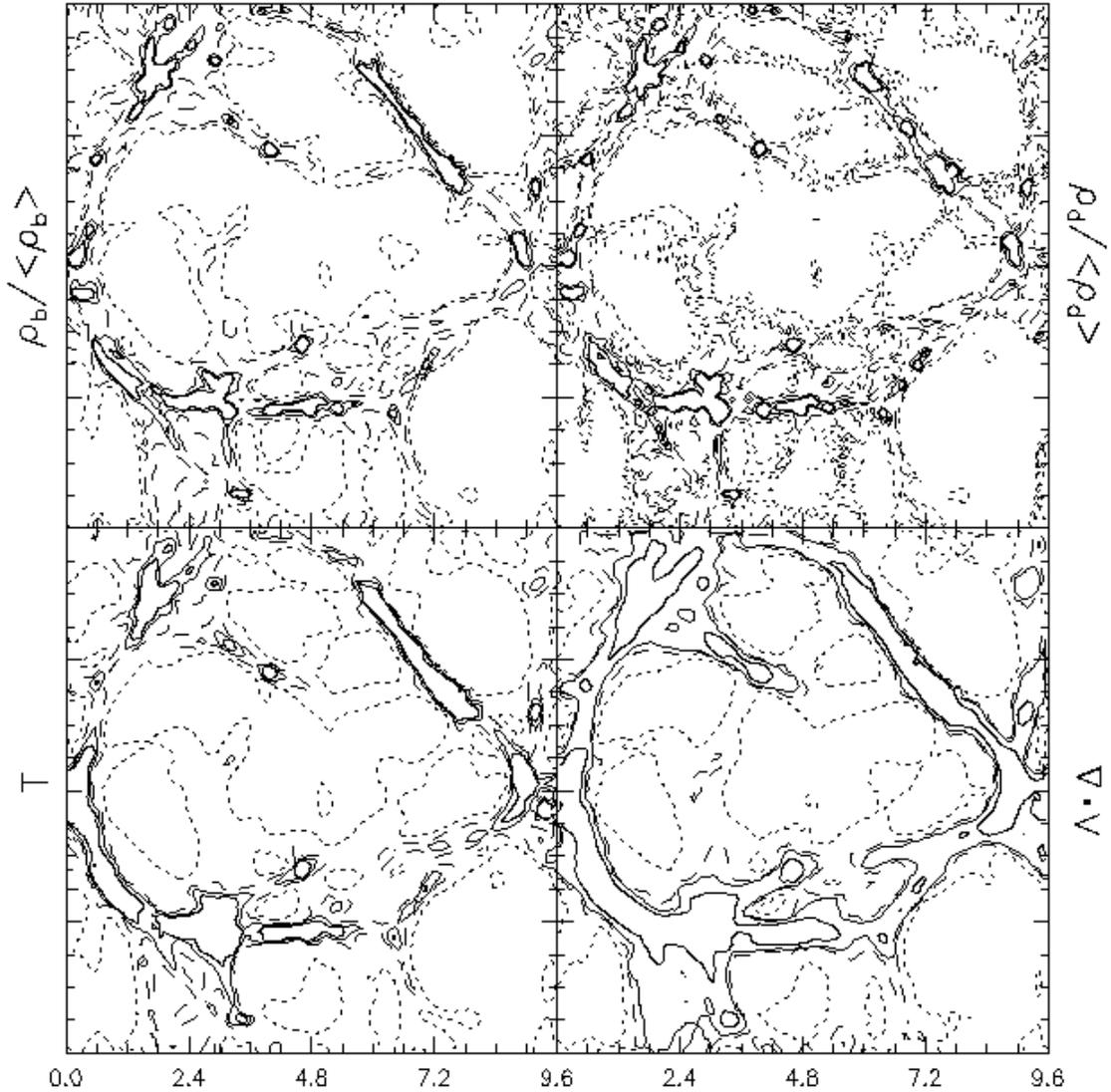}
\caption{
Two--dimensional contour plots of the baryon and dark matter overdensity
distribution, the temperature, and the divergence of the peculiar velocity
field, at $z=3$. The slice shown has a thickness of 1/16 of the box size, or
150~kpc at $z=3$. The contour levels for the baryon and dark matter densities
are set at:\ 0.5 ({\it dotted}), 1 ({\it dashed}), 3 ({\it thin solid}), and 5
({\it thick solid}). The temperature contour levels, in units of $10^3$~K,
are:\ 6 ({\it dotted}), 10 ({\it dashed}), 14 ({\it thin solid}), and 20
({\it thick solid}). The peculiar velocity divergence contours, in units of
$H_0$, are:\ 5 ({\it dotted}), 0 ({\it dashed}), -3 ({\it thin solid}),
and -15 ({\it thick solid}).
Overdense regions tend to be associated with warmer gas and a negative
peculiar velocity divergence, indicating collapse. By contrast, the underdense
regions are expanding, resulting in lower temperatures than given by
photoionization thermal equilibrium.}
\label{fig:cont_z3}
\end{figure}
In Figure \ref{fig:cont_z3}, we show two dimensional contour plots of the
baryons and the dark matter, as well as the baryonic temperature and peculiar
velocity divergence, at $z=3$. Three--dimensional representations of the
results from this and an earlier simulation (Zhang \etal 1996), are available
at our WWW site ${\rm http://zeus.ncsa.uiuc.edu:8080/LyA/minivoid.html}$,
and in Norman (1996). The plots in Figure \ref{fig:cont_z3} show a slice with
a width $1/16$ of the box size. There are obvious morphological
distinctions between the high and low density structures. The high
overdense structures (thick solid lines), are typically isolated
objects with an elongated or spheroidal baryon distribution. The
baryon distribution closely follows the dark matter
distribution. Typical thicknesses of the structures are $30-100$ kpc,
though the baryons show coherence over scales as great as a
megaparsec. Both the baryons and the dark matter show alignments in
their isolated structures over megaparsec scales,
defining a network of sheets and filaments. The intermediate density
absorbers (solid and dashed lines), are associated mainly with the
filaments or sheets themselves. The negative peculiar velocity
divergence of these systems shows that they are still undergoing
gravitational collapse. The lowest density clouds arise in the
underdense ($\rho_b/\overline{\rho}_b < 1$), minivoids
between the filaments. The underdense absorbers (dotted lines), tend
to be irregular in shape and have no preferred direction as in the
case of the filaments and sheets. They typically have temperatures
lower than the equilibrium temperature of $\sim10^4$~K and a positive
peculiar velocity divergence, indicating that these regions are
expanding with respect to the Hubble flow.  There is a general
correspondence between high (low) overdensities, high (low)
temperatures, and large (small) negative peculiar velocity
divergences.

\begin{figure}
\plotone{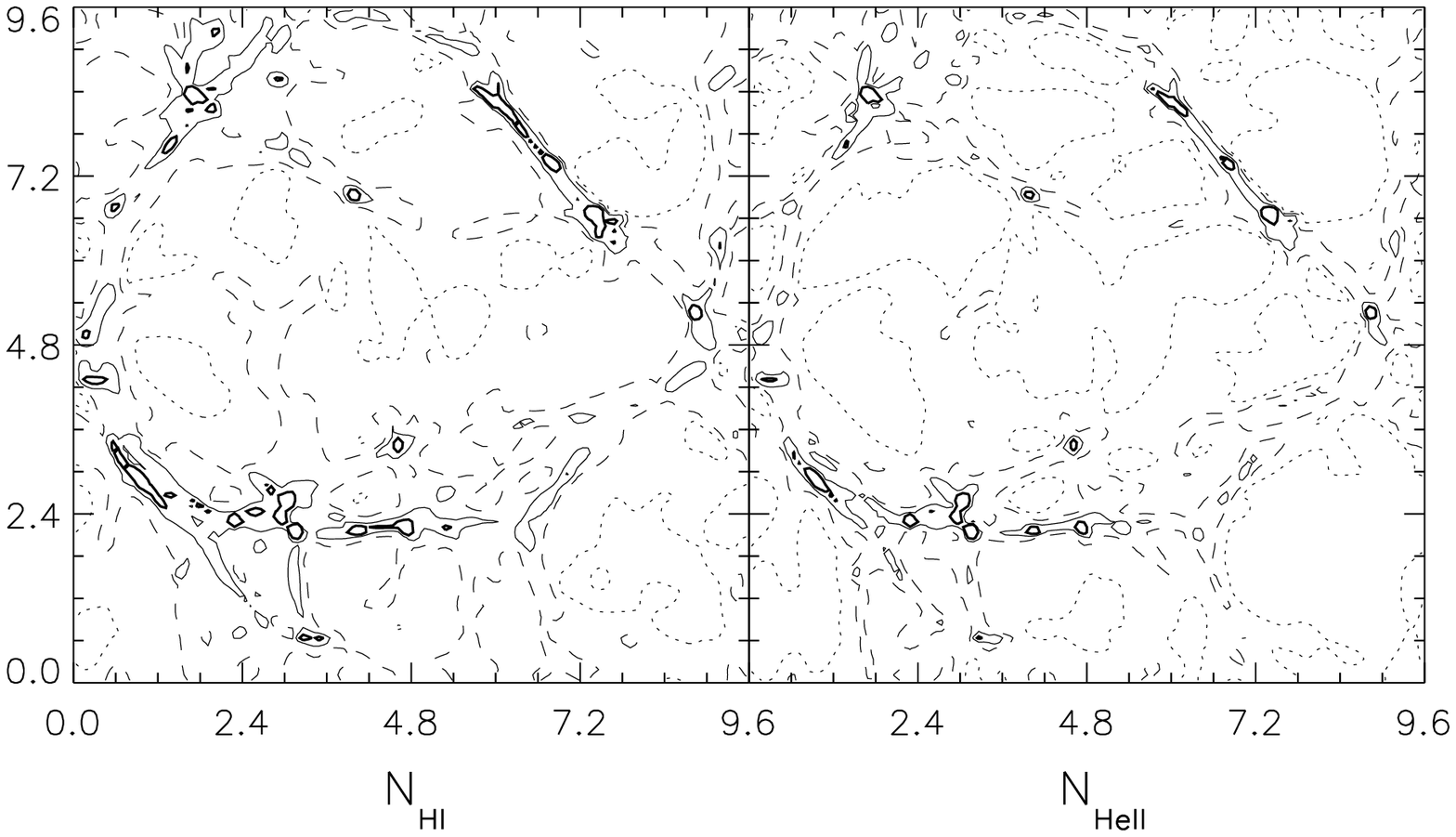}
\caption{
Contour plots of the \HI and \HeII column density distributions at $z=3$,
integrated directly through 1/16 of the box size. The contour levels
corresponding to the dotted, dashed, thin, and thick solid lines are,
respectively, $\log\NHI=12$, 13, 14, and 15, and $\log N_{\rm HeII}=14$,
15, 16, and 17. The optically thin \HI systems ($\log \NHI<13$), are associated
with the underdense regions shown in Figure 1. The saturated
lines are associated with the filaments and sheets of moderate overdensity.
The highest column densities ($\log\NHI>15$) coincide with the largest
spheroidal overdense systems at the nodes of intersecting filaments.}
\label{fig:cont_col}
\end{figure}
The corresponding \HI and \HeII column density contours are shown in
Figure \ref{fig:cont_col}. A comparison with Figure \ref{fig:cont_z3}
shows a clear correlation between column density and overdensity. The
high column density absorbers ($\NHI>10^{15}\cm2$ and
$\NHeII>10^{17}\cm2$) correspond to the highly overdense
structures ($\rho_b/\overline{\rho}_b >10$) residing mostly along and
at the intersections of filaments. The medium column density
absorbers ($\sim10^{13-14}\cm2$ for \HI and $\sim10^{15-16}\cm2$ for
\HeII) correspond to the modestly overdense filaments
($1<\rho_b/\overline{\rho}_b<5$). The column density can be coherent
over the scale of a few megaparsecs at this level. The lowest column
density absorbers ($\sim10^{12}\cm2$ for \HI and $\sim10^{14}\cm2$ for
\HeII) are associated with underdense structures
($\rho_b/\overline{\rho}_b<1$) and are located in the void regions
between the filamentary structures. They are typically a few hundred
kiloparsecs across.

\begin{figure}
\caption{
An evolutionary sequence of the baryon overdensity distribution, shown at
$z=2$, 3, 4, and 5. Superimposed is the peculiar velocity field. A region
of spheroidal collapse is visible in the mid left hand portion, while
collapse onto an oblique sheet occurs slightly to the right of center. Notice
the flow along the elongated structures leading into the pancake,
particularly at $z=3$ and $z=2$. Only moderate evolution
in the overdensity occurs in the sheets and filaments, which themselves are
overdense by only a factor of a few. Most of the universe is occupied by
underdense regions, with an associated divergent peculiar velocity field.}
\label{fig:rho_evol}
\end{figure}

An evolutionary sequence is shown in Figure \ref{fig:rho_evol} of the
baryonic overdensity over the redshift range $z=5$ to z$=2$.
Superimposed is the peculiar velocity field projected onto the plane
of the slice. As the universe evolves towards the lower redshifts, the
average density decreases due to cosmic expansion, yet the structures
remain nearly constant in morphology. There is some indication that
the filaments sharpen with time, and that the dense regions at the
intersections become more dense, but by and large the principal
structures are in place by $z=5$. The simulations show that once in
place, the sheets and filaments maintain a nearly constant overdensity
with time. By contrast, the collapse of a halo is apparent along the
lower left side of the plot.  The halo is at the center of a strongly
convergent velocity flow, and continues to accrete material from the
surrounding underdense regions and from along the filaments throughout
the simulation. A pancake is visible slightly offset to the right from
the center of the figure, at an oblique angle. The continual collapse
of material onto the pancake is indicated by the velocity field, which is
nearly perpendicular to the pancake surface. Notice the flow along the
elongated structures leading into the pancake as well, as is
especially clear in the figure at $z=3$ and $z=2$. The general pattern
is one of outflow from the minivoids, compression into sheets at the
boundaries of the voids, and a resulting flow along the sheets toward
their intersections. The underdense regions themselves are not uniform,
but reveal small--scale mottling and striations. All these structures
give rise to discrete absorption features in our synthesized spectra.

\begin{figure}
\plottwo{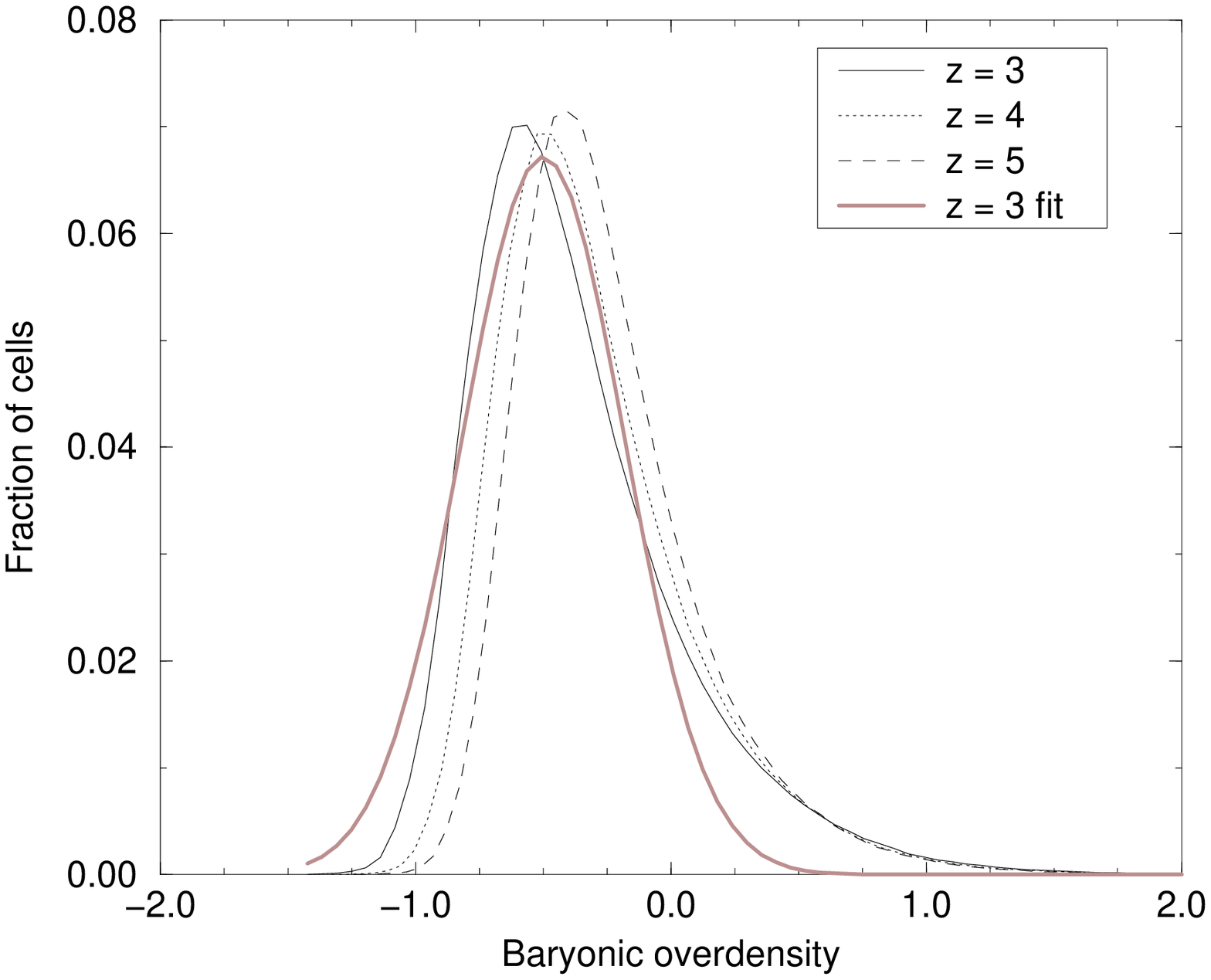}{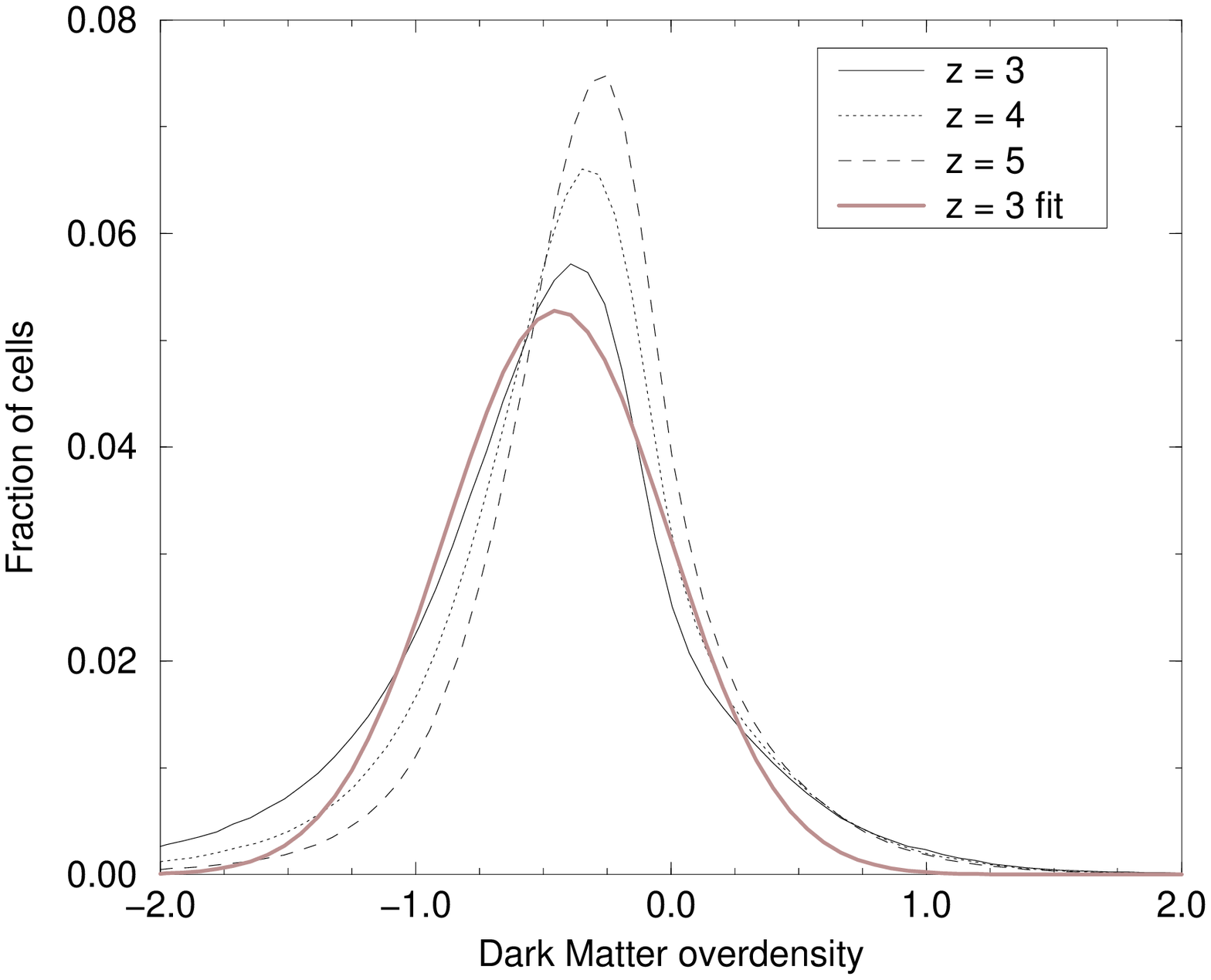}
\caption{
The distribution of baryon overdensity, at $z=3$, 4, and 5. Consistent
with the slow evolution in Figure 3, the distributions
are nearly identical, with a drift of the mode toward lower values as
the underdense regions continue to vacate. The long tail toward higher
overdensities accounts for the higher column density \HI systems.
Also shown is the best fit to a lognormal distribution at $z=3$.}
\label{fig:odb_distr}
\caption{
The distribution of the dark matter overdensity, at $z=3$, 4, and 5. Most
of the distribution is well--fit by a lognormal distribution, although
tails in the distribution occur at the low and high density extremes.
Also shown is the best fit to a lognormal distribution at $z=3$.}
\label{fig:odDM_distr}
\end{figure}
The slow evolution of the baryon density is reflected by the
distribution of the baryon overdensity, shown in Figure
\ref{fig:odb_distr}. The mode of the distribution shifts to lower
values as the underdense regions continue to vacate, resulting in a
broadening of the overall distribution. For comparison, we show the
evolution of the dark matter overdensity distribution in Figure
\ref{fig:odDM_distr}. Theoretical considerations suggest that the
density distribution should be approximately lognormal (Coles \& Jones
1991; Colombi 1994). We find that a lognormal distribution fits both
the baryon and the dark matter overdensity distributions moderately
well, as was similarly found for the cell occupation densities in the
simulations of Coles, Melott, \& Shandarin (1993). Both distributions,
however, have pronounced high density tails, so that the probability
distribution of the rarer overdense structures is distinctly not
lognormal. At low densities, the baryon density distribution cuts off
more sharply than a lognormal distribution, while the dark matter
distribution shows a broad wing that is also not lognormal.

\begin{figure}
\caption{
An evolutionary sequence of the \HI column density, at $z=2$, 3, 4, and 5.
In contrast to the baryonic overdensity, the \HI column density levels
break up with time, and the universe becomes more transparent as it evolves.}
\label{fig:NHI_evol}
\end{figure}
The constancy of the baryonic overdensity in the sheets is an expected
consequence of pancake collapse. The computations of Meiksin (1994) for the
collapse of photoionized gas into slabs show that a weak counterflow
develops after the formation of a caustic in the dark matter
distribution and the subsequent diminishing of the gravitational
potential of the slab. This outflow meets the larger inflow in an
accretion shock, the position of which moves outward from the slab
center. The result is a baryonic density within the slab that
decreases nearly in proportion to the cosmic average density. The
decrease in baryon density is reflected by a rapid decrease in the
\ion{H}{1} column density of the slab. In Figure \ref{fig:NHI_evol}, we
show the evolution of the column density in the simulation. A consequence
of the evolution is that the universe becomes more transparent with time.
The number of systems detected along a given line--of--sight above a fixed
column density threshold decreases toward decreasing redshift, at a
rate consistent with observations (Zhang \etal 1997). The contours,
{\it at the same fixed levels}, become less connected and the sheets
and filaments become less extended in space. Despite this general
contraction, the {\it total} number of lines actually increases per
unit redshift at lower redshifts as the comoving scale increases for a
fixed redshift interval (Zhang \etal 1997, Figure 12).

\subsection{Physical State of the Clouds}
\label{subsec:state}

\subsubsection{Correlations with HI Column Density}
\label{subsubsec:corrHI}

\begin{figure}
\plottwo{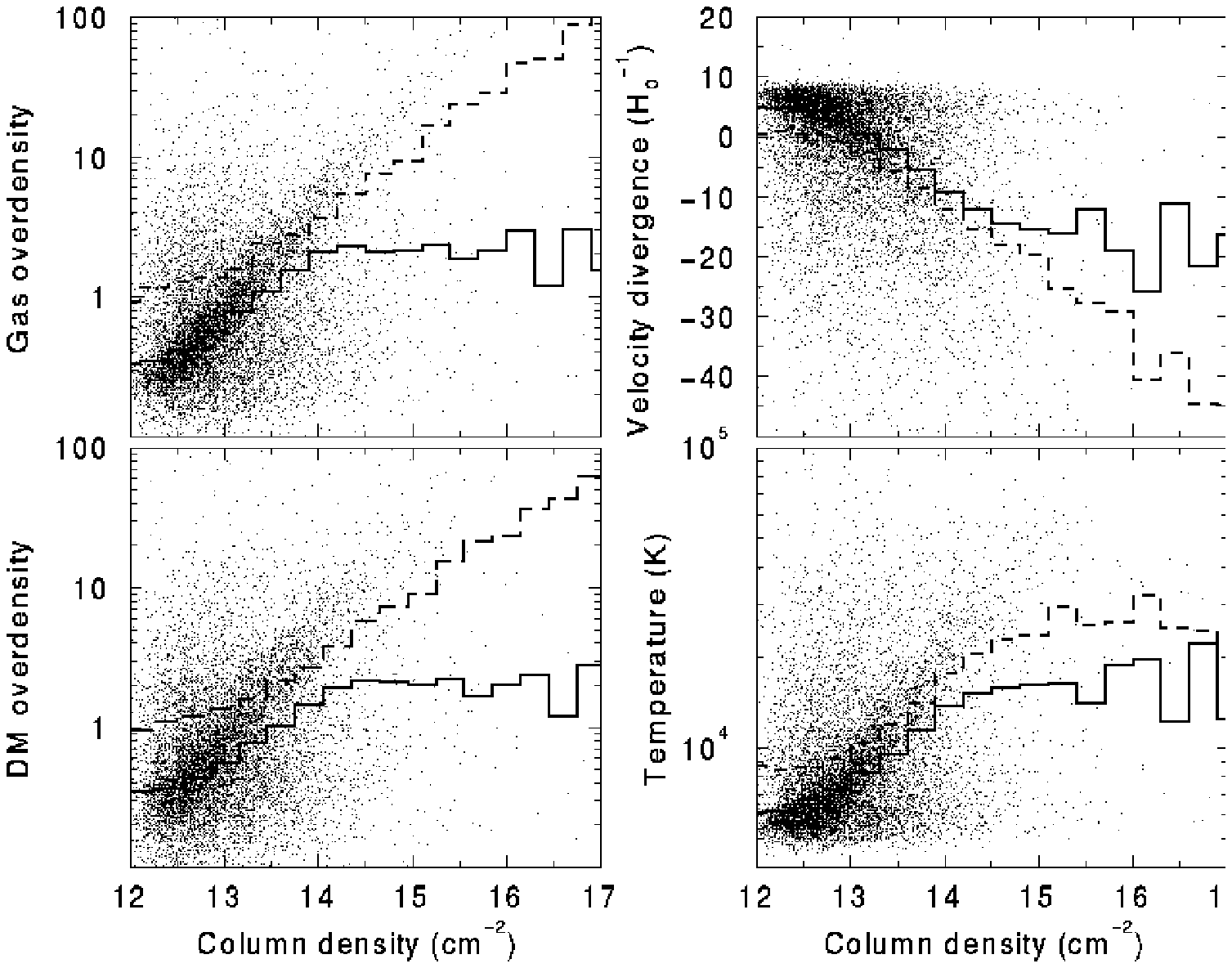}{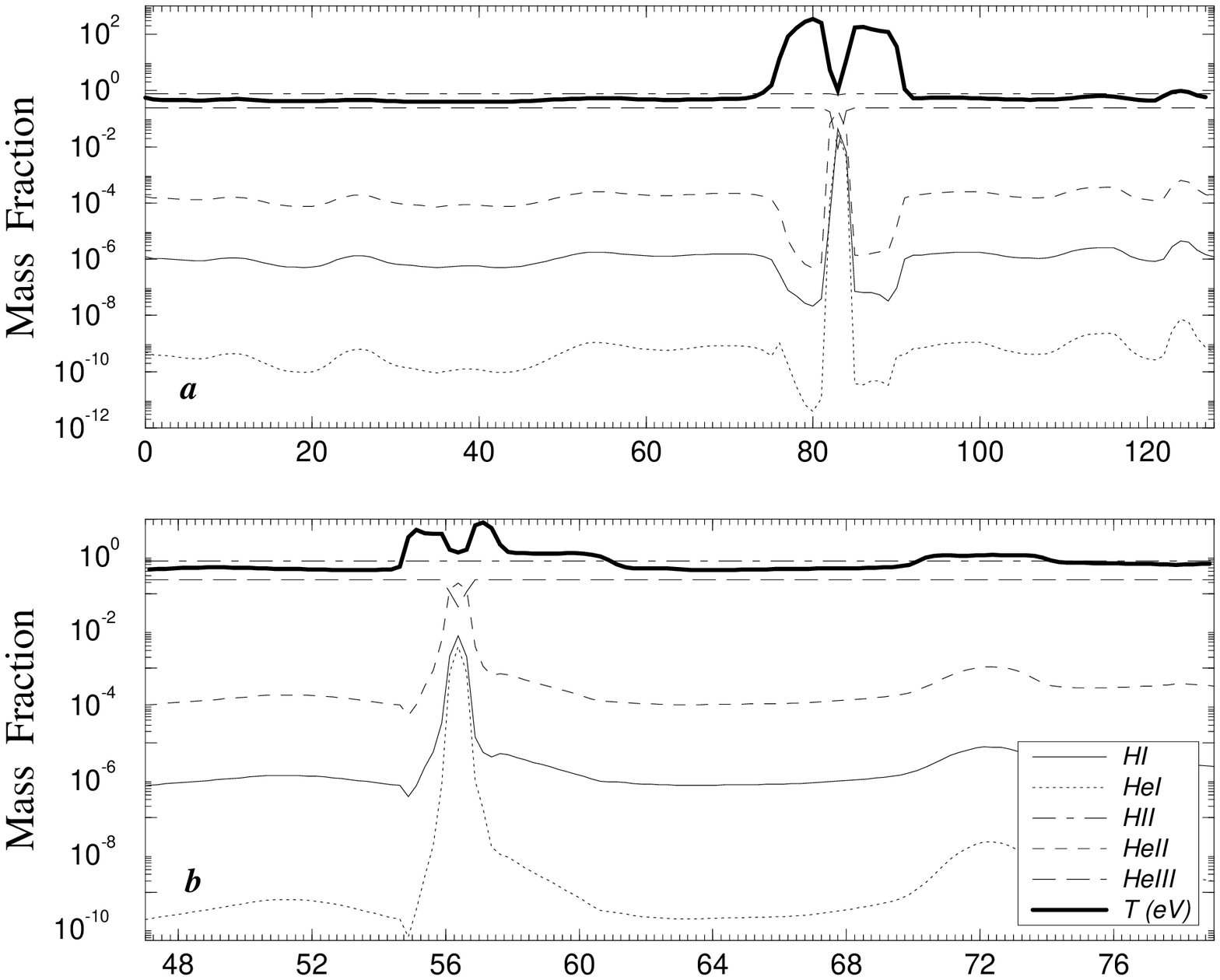}
\caption{
Scatter plots showing the correlations between cloud baryon and
dark matter overdensities, divergence of the peculiar velocity, and temperature
with \HI column density, at $z=3$. Also shown are the median ({\it solid line})
and mean ({\it dashed line}) of the distributions. Although there is
substantial scatter, the internal physical properties of the absorbers
correlate extremely well with the \HI column density.}
\label{fig:corr_NHI}
\caption{
Line plots across a single cell at $z=3$ through the
densest structures in the (a)\ top and (b)\ subgrids. The $x-$axis is
the cell number. The axis for the subgrid is renumbered to show that
the subgrid resolution is four times that of the top grid. The
$y-$axis is the mass fraction of each species, relative to the total
baryonic mass, and the temperature is in units of eV. (Note that the
two plots correspond to different lines through the box since the
densest structure in the top grid is not covered by the subgrid.)}
\label{fig:stat_mul}
\end{figure}
A strong correlation exists not only between baryon density and \HI
column density, but between temperature, peculiar velocity divergence,
and the ratio of dark matter to baryonic overdensities as well. In
Figure \ref{fig:corr_NHI} we show scatter plots of these quantities,
along with the average and median dependences on $\NHI$.
(Also see Zhang \etal 1997; Meiksin 1997).
Similar correlations were found for a $\Lambda$CDM model (Miralda-Escud\'e
\etal 1996). The temperature of the
baryons varies over the range of about 5000 K in expanding regions to
over $10^6$ K in strongly collapsing regions. Most of the gas mass is in
photoionization thermal equilibrium with the ionizing background, and
so has a characteristic temperature of $15-30\times10^3\kel$, as shown
in Figure \ref{fig:cont_z3}. In Figure \ref{fig:stat_mul}a we show a
line plot across the densest structure on the top grid of the 9.6 Mpc
simulation at redshift $z=3$. The temperature in most of the cells is
elevated to about $1\eV$ due to photoionization heating by the UV
radiation background. The densest structures on the grid have a
typical caustic--like shape, i.e. the outskirts of the collapsing gas
are shock--heated to more than $100\eV$ due to the infalling gas,
while the center of the structure is radiatively--cooled (mostly by
hydrogen line cooling) to the hydrogen recombination determined
temperature of $\sim 1\eV$. Similar structures were found in the
simulations of Cen \etal (1994). The gas is completely ionized except
in the center--most regions of the densest structures. Typical
fractional abundances of \HI and \HeII are roughly $10^{-6}$ and
$10^{-4}$ respectively. However, at the centers of the densest
structures, hydrogen and helium have mostly recombined to their
neutral forms. Figure \ref{fig:stat_mul}b shows the corresponding line
plot across the densest cell in the sub grid simulation. Because the
subgrid is positioned over the most {\it underdense} void region of
the top grid, the temperatures of the densest structures found on the
subgrid are typically lower than those on the top grid.

\begin{figure}
\plottwo{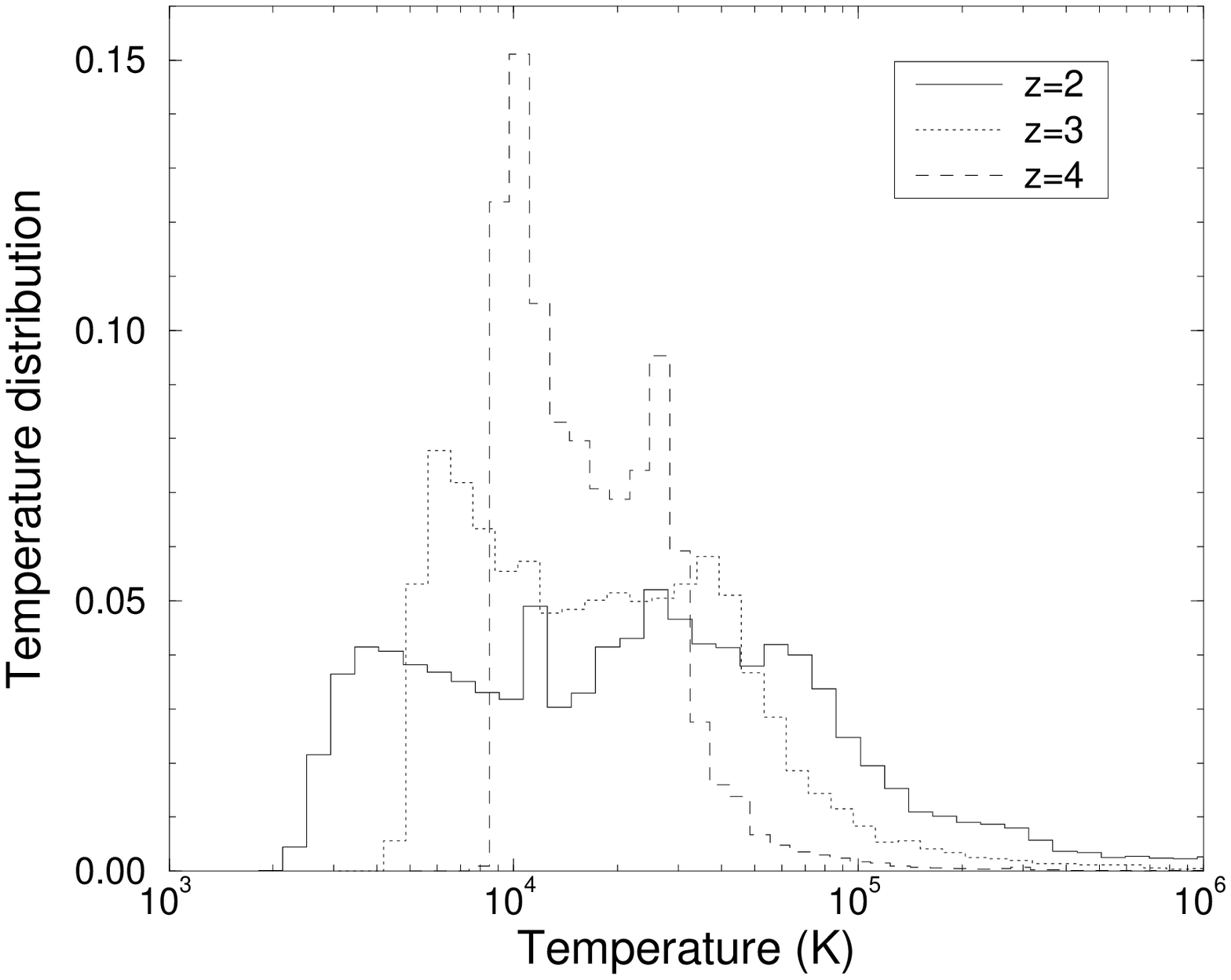}{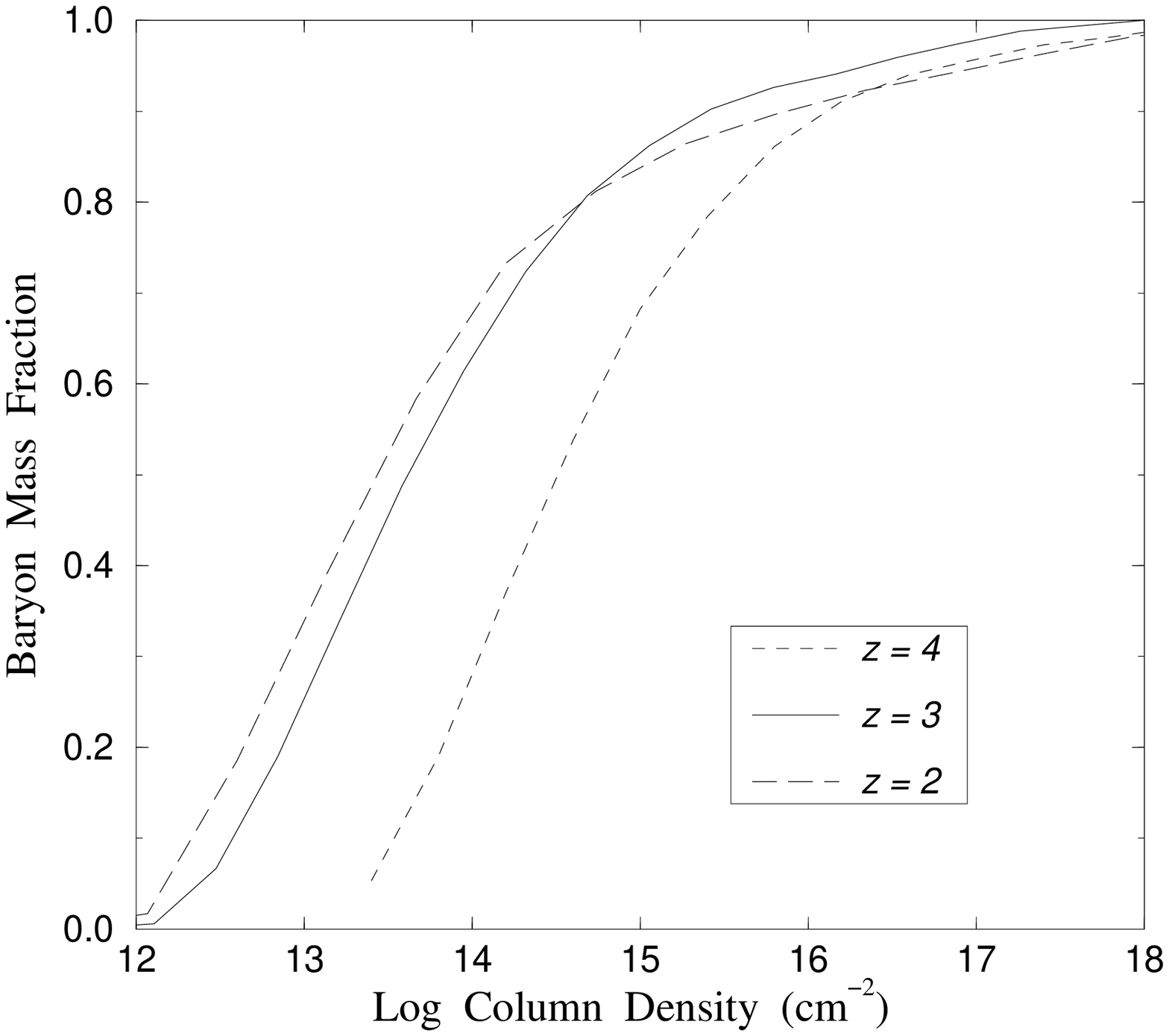}
\caption{
Gas temperature distribution, weighted by baryon density, at $z=2$
({\it solid line}), 3 ({\it dotted line}), and 4 ({\it dashed line}).
While at $z=4$ most of the gas is near the temperature corresponding
to photoionization thermal equilibrium, the distribution flattens with
time as massive regions continue to form and accrete material and the
underdense regions continue to cool due to adiabatic expansion.}
\label{fig:temp_distr}
\caption{
Cumulative distributions of the baryons below a given \HI column density
at $z=2$, 3, and 4. Most of the baryons in the IGM lie within structures that
give rise to discrete absorption features with column densities
$12.5 < \log\NHI < 14.5$ at $z=3$. Fewer than 5\% of the baryons remain in
unresolved systems with $\log\NHI<12.5$ to be detected as a smoothly
distributed uniform component.}
\label{fig:bfrac}
\end{figure}

The high temperatures ($T\gg 10^4\kel$) found are readily
explained by gravitational infall. Gas falling into a deep potential
well will be shock--heated to the effective virial temperature of the
potential well, only to relax to the photoionization equilibrium value
downstream. Low temperatures ($T\ll 10^4\kel$), however, are
also found. These occur in low density regions, where the density of
the baryons is too low for the gas to maintain equilibrium at the
photoionization temperature against the expansion of the gas. This
effect was discussed by Meiksin (1994), where it was shown that when
the gas density falls below $n_{\rm H}\simeq10^{-4}\pcc$ for $z>2$ in
an otherwise quiescient region, the expansion of the gas will drive it
out of photoionization thermal equilibrium, resulting in a temperature
below the equilibrium value. The reason is that the rate of energy
deposition per baryon by photoionization is proportional to the \HI
and \HeII fractions, which decline with decreasing total baryon
density. Since the energy of the photoejected electrons must be shared
with all the baryons, the rate of energy deposition per particle by
the radiation field is diminished in lower density regions. For
densities $n_{\rm H}<10^{-4}\pcc$, the energy deposition rate lags
behind the cooling rate due to cosmic expansion. A consequence of this
effect is that the temperature of the low column density systems will
be sensitive to the assumed average baryon density and to the ionization
history of the IGM. As the universe evolves, a greater fraction of
the baryons are heated to high temperatures by compression as more
massive systems continue to collapse. At the same time, the baryons
in the underdense regions continue to cool through adiabatic expansion.
The result, shown in Figure \ref{fig:temp_distr}, is a progressive
flattening of the temperature distribution with time.

\subsubsection{Characteristic Cloud Sizes and Masses}
\label{subsubsec:size}

We may use the correlations in Figure \ref{fig:corr_NHI} to derive several key
properties of the absorbers. Because of the correlation between density and
column density, the neutral hydrogen fraction will be correlated with the
column density as well. This relation may be used to derive the distribution
of baryons within the forest as a function of column density (see the
following section). A cloud with an internal density $\rho_b$ that is optically
thin at the Lyman edge and in ionization equilibrium with the
metagalactic radiation field will have a neutral hydrogen density of
\begin{equation}
n_{\rm HI}\simeq7\times10^{-15}\,{\rm cm^{-3}}\,\left(\frac{\rho_b}{\bar\rho_b}
\right)^2(1+z)^6T_4^{-0.75}\Gamma_{\rm HI,-12}^{-1}, \label{eq:nHI}
\end{equation}
where $\bar\rho_b$ is the cosmic mean baryon density, here taken to correspond
to $\Omega_bh_{50}^2=0.06$. The temperature factor is due to the temperature
dependence of the radiatiave recombination rate. Figure \ref{fig:corr_NHI}
shows that, in the column
density range $12.5<\log\NHI<14.5$, the internal cloud density varies with
column density as $\rho_b/\bar\rho_b\approx N_{\rm HI, 13}^{1/2}$ at $z=3$,
where $N_{\rm HI, 13}$ is the \HI column density in units of $10^{13}\cm2$.

A direct consequence of the square--root dependence of the cloud
density on \HI column density over the range $12.5<\log\NHI<14.5$ is
that the clouds in this range must all have nearly the same
characteristic dimension. Taking the line--of--sight total scale length
of the absorbers to be $\ell\equiv\NHI/n_{\rm HI}$, we find from
eq.(\ref{eq:nHI}) and $\rho_b/\bar\rho_b\approx N_{\rm HI, 13}^{1/2}$
that $\ell\approx100-150\kpc$, with a weak dependence on column density
through the cloud temperature. The associated characteristic cloud baryonic
masses are $M_c\equiv\rho_b(\ell)^3\approx0.3-3\times10^9\msun$, or
greater allowing for the elongation into filaments. The inferred sizes are
close to the typical filament widths shown in Figure \ref{fig:cont_z3},
and are consistent with the results of the double line--of--sight analysis
performed by Charlton \etal (1997). It is also close to the scale height
expected on dimensional grounds for photoionized gas in hydrostatic equilibrium
within the potential well of a moderate dark matter overdensity. For a
given dark matter overdensity $\rho_{\rm DM}/ \bar\rho_{\rm DM}(>1)$, the
equation of hydrostatic equilibrium for an isothermal gas,
$c_s^2\nabla\log\rho_b=-\nabla\phi$, where $c_s$ is the isothermal
sound speed, and introducing a scaleheight $r$, becomes
$c_s^2/r\approx(1/2)(\rho_{\rm DM}/\bar\rho_{\rm DM}) H(z)^2r$, where we have
expressed the gravitational potential (up to an additive constant), as
$(1/4)(\rho_{\rm DM}/\bar\rho_{\rm DM})H(z)^2r^2$. We then obtain for a
typical total thickness
\begin{equation}
\ell\approx2r\approx2^{3/2}\left(\frac{\rho_{\rm DM}}{\bar\rho_{\rm DM}}\right)
^{-1/2}\frac{c_s}{H(z)}\approx30-70\kpc, \label{eq:lscale}
\end{equation}
for $\rho_{\rm DM}/\bar\rho_{\rm DM}\approx1-5$.
From eq.(\ref{eq:nHI}), and taking the overdensity in baryons to be comparable
to that in the dark matter, the typical column densities of the moderately
overdense systems is then $10^{13}<\NHI<10^{15}\cm2$, consistent with Figure
\ref{fig:corr_NHI}. We may compare this with the minihalo model, for which the
baryonic overdensities of the halo cores will reach the higher values of
$200-1000$ due to spheroidal collapse and cooling (Meiksin 1994). The typical
sizes of these systems will be a few
kiloparsecs, and the corresponding column densities and baryonic masses will be
$\NHI\approx10^{16}-10^{17}\cm2$ and $M_c\approx10^8-10^9\msun$.
These systems correspond to the nodes of the filaments in
Figure \ref{fig:cont_z3}, though their internal structure is likely not fully
resolved in the simulation. We anticipate studying
these structures in greater detail using future higher resolution simulations.

Some of the statistical properties of the \Lya forest have recently been shown
to be derivable from a few basic physical assumptions. Bi \& Davidsen (1997)
are able to reproduce results in good agreement with the observed cloud
properties over the column density range $10^{13}<\NHI<10^{15}\cm2$ starting
with a lognormal distribution for the baryon density, and a suitably adjusted
cloud scale length. The differences between the true distributions and a
lognormal (Figures \ref{fig:odb_distr} and \ref{fig:odDM_distr}), though,
will necessarily introduce a bias in the normalization of the column density
distribution and the inferred value of the cosmological baryon density.
Hui, Gnedin, \& Zhang (1996) are able to obtain a very good match to the
column density distribution and Doppler parameters in the simulations using a
semianalytic approach based on the truncated Zeldovich approximation (Coles
\etal 1993). The agreement with
the fully self--consistent hydrodynamical calculations offers the promise
of using semianalytic methods to infer the effect of variations in the
cosmological models on the dominant statistical properties of the \Lya forest.

\subsection{Ionization State of the Clouds}
\label{subsec:cloudionization}

Recently it has become possible to place constraints on the ionization
parameters of the \Lya forest systems by measuring the column
densities of various ionization stages of carbon and silicon.  Using
the Keck HIRES, and assuming a cloud temperature determined by the
balance of photoeletric heating and radiative losses, Songaila \&
Cowie (1996) inferred a typical ionization parameter for clouds at
$z\approx3$ with $\log\NHI>15$ of $U\approx0.02$, with most clouds
falling in the range $0.003<U<0.03$. The ionization parameter
is defined to be $U=n_\gamma/n_{\rm H}$, where $n_\gamma$ is the
number density of hydrogen ionizing photons in the ambient radiation
field, and $n_{\rm H}$ is the total number density of hydrogen atoms. The
best fitting models require a large break in the UV background at the
\ion{He}{2} photoelectric edge, in keeping with the break required by
our simulation to reproduce the measurements of the intergalactic
\ion{He}{2} opacity (Zhang \etal 1997). By comparing the pair of
column density ratios \ion{C}{2}:\ion{C}{4} and \ion{Si}{4}:\ion{C}{4},
they concluded that silicon must be enriched relative to carbon by a factor
of a few to several above the solar abundance ratio, consistent with
evidence for a high abundance of $\alpha$--processed material found in
some damped \Lya systems at high redshift (Lu \etal 1997).

The ionization parameter scaling with the cosmic mean density and the
intensity of the UV ionizing background differs by a factor of
$\Omega_b$ from $b_{\rm ion}$, giving $U\propto\Omega_b/b_{\rm
ion}\propto(n_\gamma/ b_{\rm ion})^{1/2}$. This breaks the degeneracy
between the baryon density and the intensity of the radiation field in
$b_{\rm ion}$ alone. Combining the normalization requirement of the mean
intergalactic \ion{H}{1} opacity, which fixes $b_{\rm ion}$, with the metal
absorption line ratios, which fixes $U$, it is possible to constrain
$\Omega_b$ and $n_\gamma$ (or $\Gamma_{\rm HI,-12}$) separately. This is
somewhat complicated by the spectral shape of the UV background, in particular
the size of the spectral break between the \ion{H}{1} and \ion{He}{2}
photoelectric edges. Measurements of the mean \ion{He}{2} \Lya
opacity, however, provide a stringent constraint on the size of the
break (Jakobsen et al. 1994; Madau \& Meiksin 1994). Based on our
spectral analysis (Zhang \etal 1997), we require $b_{\rm ion}=0.004-0.006$
for \ion{H}{1} at $z=3$, and a spectral break between
the \ion{H}{1} and \ion{He}{2} edges corresponding to $\Gamma_{\rm
HI}/\Gamma_{\rm HeII}=250-400$.

We may use the simulation results to assess the values of $\Omega_b$
and $n_\gamma$ by comparing the cloud properties we find in the
simulation with those inferred by Songaila \& Cowie (1996). The principal
constraint on the low ionization parameter inferred from the
observations arises from the high \ion{Si}{4}:\ion{C}{4} ratio
measured. The \ion{C}{2}:\ion{C}{4} and \ion{C}{4}:\ion{H}{1} data alone
are consistent with a higher value of $U\approx0.2$ and 1\% solar abundances;
however, for this value of $U$, \ion{Si}{4} would be undetectable
(Bergeron \& Stasi\'nska 1986). Of 18 systems in the column density
range $15<\log\NHI<17$ identified by Songaila \& Cowie in two
lines-of-sight, almost all (17) show \ion{C}{4} absorption, half (9)
show \ion{Si}{4} absorption, and a third (6) have
\ion{Si}{4}:\ion{C}{4}$>0.1$. It is the high incidence of this last
ratio we must explain.

Haardt \& Madau (1997) have computed the UV background produced by QSO
sources with an intrinsic spectral index of $\alpha_Q=1.8$ and 2. (For
the simulation, we used their earlier spectra based on $\alpha_Q=1.5$.)
These spectra are based on the steep indices measured by Zheng \etal (1996)
using {\it HST} data. We compare our results assuming the new spectra.
At $z=3-3.5$, the spectra give $\Gamma_{\rm HI}/\Gamma_{\rm HeII}\approx
210-220$ for $\alpha_Q=1.8$, with $\Gamma_{\rm HI,-12}\approx
0.6-0.8$, while $\Gamma_{\rm HI}/ \Gamma_{\rm
HeII}\approx330-340$ and $\Gamma_{\rm HI}\approx 0.5-0.7$ for
$\alpha_Q=2$. The density of \ion{H}{1} ionizing photons is
$n_\gamma\approx0.5-1\times10^{-5}\pcc$ for both cases. An ionization
parameter of $U=0.02$ then requires an internal cloud total hydrogen
density of $n_{\rm H}\approx2.5-5\times10^{-4}\pcc$, corresponding to
a baryonic overdensity of $\rho_b/\bar\rho_b\approx25-65$. We find in
the simulation at $z=3$ that for the systems with $15<\log\NHI<17$,
the fractions of systems with $\rho/\bar\rho_b>25$, 50, and 100, are
30\%, 20\% and 5\%, respectively. Alternatively, the value of the mean
cosmic baryon density could be increased. Combining with the limit
above on $b_{\rm ion}$ from the \ion{H}{1} spectral analysis, we find
that we may obtain a comparable number of dense systems to that
observed, but only if (1)\ there is a large break in the UV background
at the \ion{He}{2} photoelectric edge of $\Gamma_{\rm HI}/\Gamma_{\rm
HeII}\approx 250-400$, (2)\ the abundance ratio of Si to C is
increased by a factor of $\sim3$ above solar (both in agreement with
the findings of Songaila \& Cowie), and (3)\ the photoionization rate
of \ion{H}{1} is $\Gamma_{\rm HI,-12}\lsim1$. Our best estimates for
the required cosmological mean baryon density and \ion{H}{1}
photoionization rate at $z\approx3-3.5$ are $0.03\lsim\Omega_b\lsim0.08$
($h_{50}=1$) and $0.3\lsim\Gamma_{\rm HI,-12}\lsim1$, respectively. The
constraints on the UV photoionizing background are in excellent agreement with
the field produced by QSOs with a steep intrinsic spectral index of
$\alpha_Q\lsim2$. We will make a more complete comparison elsewhere.

Rauch, Haehnelt, \& Steinmetz (1996) recently performed an analysis of the
expected metal ion ratios for a simulation designed to study galaxy formation,
finding good agreement with the observed metal ratios, provided the Si to C
ratio was higher than solar. Since their simulation was deliberately chosen to
include several protogalactic clumps, it is unclear how representative the
absorber properties they find are of the cosmological distribution. For
example, they find $\log\NHI=14$ systems to be associated with a baryonic
overdensity of 10--20 at $z=3$ (their Figure 6), a factor of 3--4 higher than
we obtain. This would result in a lower ionization parameter by the same
factor. A simulation closer to ours is that of Hellsten \etal (1997). They
claim to find a reasonable match to the Songaila \& Cowie data after increasing
the break at the \ion{He}{2} photoelectric edge a factor of 10 over that of the
Haardt \& Madau spectrum, and assuming a high Si to C ratio.

\subsection{The Baryonic Mass in the \Lya Forest}
\label{subsec:baryons}

The contribution of the baryons in the Ly$\alpha$ forest to the closure
density is given by,
\begin{equation}
\Omega_{\rm Ly\alpha}=\frac{1.4m_{\rm H}}{\rho_{\rm crit}}\frac{H_0}{c}
\int dN_{\rm HI}\,\frac{N_{\rm HI}}{f_{\rm HI}}\frac{\partial^2 N}
{\partial N_{\rm HI}\partial z}(1+z)^{5/2}, \label{eq:OLya}
\end{equation}
where $f_{\rm HI}\equiv n_{\rm HI}/n_{\rm H}$ is the mean fraction of neutral
hydrogen in a cloud of a given column density. Using eq. (\ref{eq:nHI}) and
the relation found between cloud overdensity and \HI column density at $z=3$
above, $\rho_b/\bar\rho_b\approx N_{\rm HI, 13}^{1/2}$, we obtain
$f_{\rm HI}\approx4\times10^{-6}N_{\rm HI, 13}^{1/2}T_4^{-0.75}$ for clouds
with $12.5 < \log\NHI<14.5$. The \HI column density
distribution varies nearly like $\NHI^{-1.5}$. We then obtain
\begin{equation}
\Omega_{\rm Ly\alpha}\propto
\log\left(\frac{N_{\rm HI, max}}{N_{\rm HI, min}}\right). \label{eq:OLyal}
\end{equation}
Most of the baryons lie in the column density range $12.5<\log N_{\rm HI}
<14.5$, distributed equally per decade in column density.
A more quantitative comparison with the distribution
of the baryons may be made by allowing for the temperature dependence of the
clouds on column density, as given in Figure \ref{fig:corr_NHI}. Taking
$T_4\simeq0.8\,N_{\rm HI, 13}^{1/4}$, so that $f_{\rm HI}\approx5\times10^{-6}
\NHI^{5/16}$, and using the column density
distribution fit over the range $2\times10^{12}<\NHI<10^{14}\cm2$ of
$\partial^2 N/ \partial N_{\rm HI, 13}\partial z\approx6
N_{\rm HI, 13}^{-1.4}(1+z)^{2.5}$ (Zhang \etal 1997), in close agreement with
Hu \etal (1995), we obtain
\begin{equation}
\Omega_{\rm Ly\alpha}\simeq3.6\times10^{-6}N_{\rm HI, max}^{0.29}
\left[1-\left(\frac{N_{\rm HI, min}}{N_{\rm HI, max}}\right)^{0.29}\right].
\label{eq:Olyan}
\end{equation}
The contribution from clouds in the range $12.5<\log\NHI<14$ is then
$\Omega_{\rm Ly\alpha}\approx0.03$, or 50\% of the baryons in the simulation.
The direct computation shown in Figure \ref{fig:bfrac} yields a comparable
value for the baryonic mass in clouds with column densities in this range.
Because we do not reproduce the column density distribution for
$\log\NHI\gsim16.5$ (Zhang \etal 1997), the distribution of the baryons among
the high column density systems is uncertain. In particular, a larger box
simulation may show a greater concentration of baryons in the highest
column density systems, especially the Damped \Lya Absorbers.
Only a tiny fraction of the baryons, less than 5\%, remains in a diffuse
component unresolved as low column density discrete systems. Even this small
component would possibly be shown to be discrete in a higher resolution
simulation. Virtually the entire intergalactic medium has fragmented.

Meiksin \& Madau (1993) and Press \& Rybicki (1993) had previously
suggested that most of the baryons may reside in the \Lya forest,
provided the clouds were large, with sizes of $\sim100$ kpc. Meiksin \&
Madau argued that this would be a natural means of reconciling the
predictions of Big Bang nucleosynthesis for the density of the IGM and
Gunn--Peterson constraints on the \HI density of a smooth diffuse IGM
component with a metagalactic UV radiation field dominated by QSO
sources. In their constant expansion rate model, for which the clouds
have a size of $100-200$~kpc at $z=2.5$, Meiksin \& Madau estimated
$\Omega_{\rm Ly\alpha}\simeq0.07$, in very good agreement with the results
presented here.

\subsection{Origin of Spectral Features}
\label{subsec:spec}

\begin{figure}
\plottwo{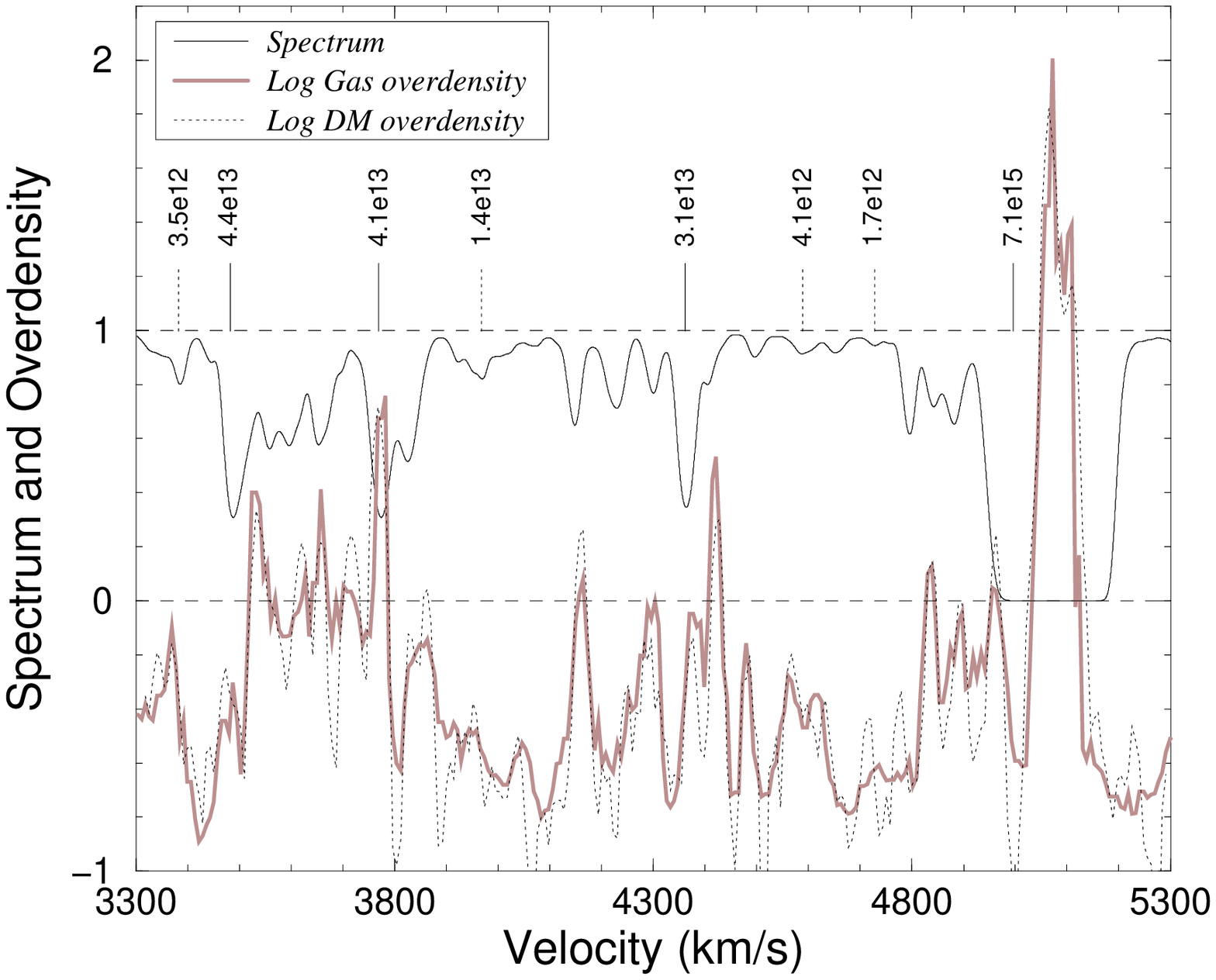}{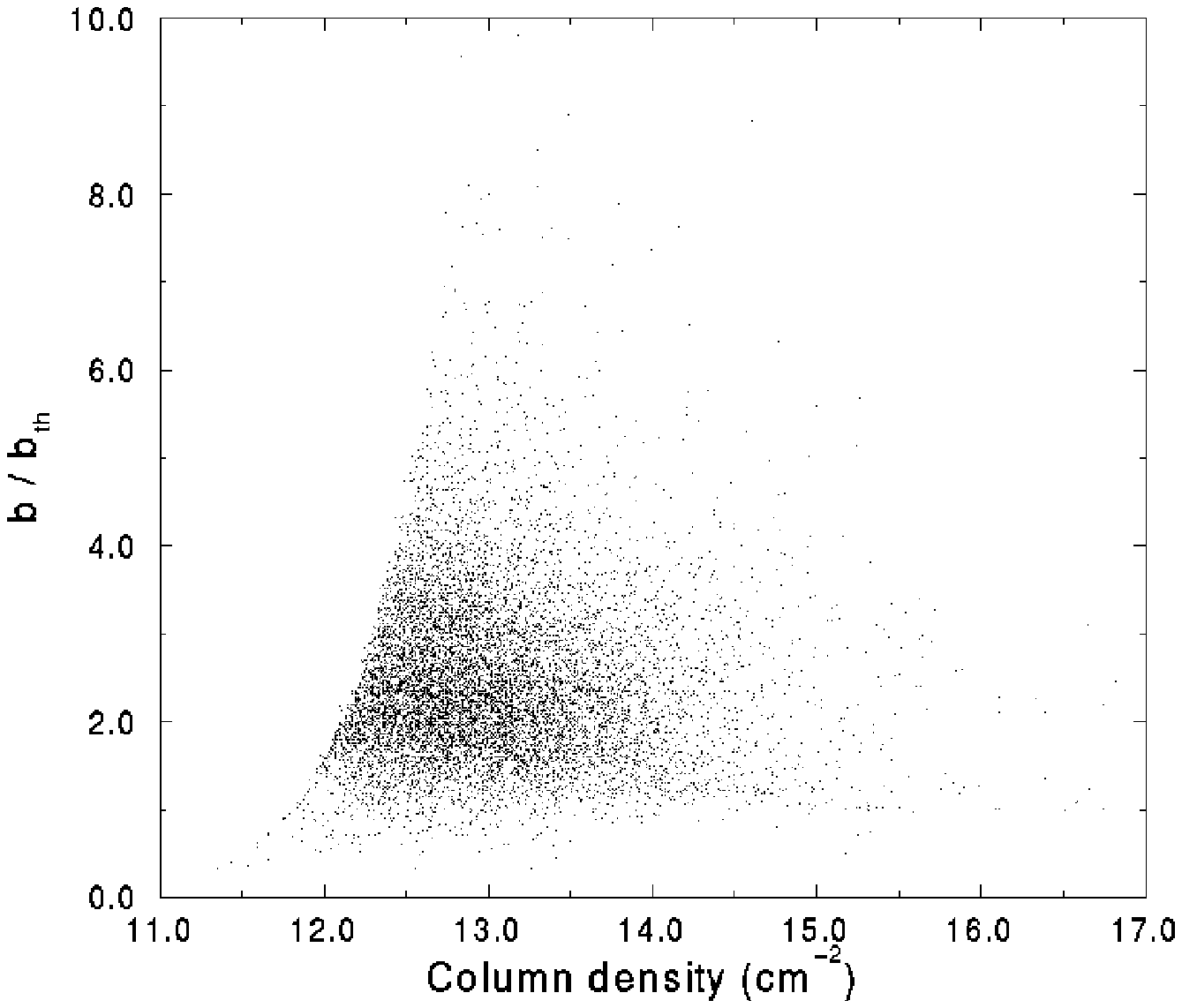}
\caption{
A sample of the absorption spectrum along a line--of--sight through the
box at $z=3$, as a function of velocity in the rest frame of the box.
Also shown are the baryonic and dark matter overdensities
that give rise to the absorption features. The densities are plotted
according to their positions in the comoving frame, while the absorption
features include the effects of the gas peculiar velocity. Systemic
offsets of up to $100\kms$ occur between a density feature in the gas
and the corresponding spectral feature. The baryon fluctuations in general
closely follow the dark matter fluctuations, even when both fluctuations
are underdense relative to the cosmic mean. These regions give rise
to optically thin absorption features.}
\label{fig:specHI_od}
\caption{
A scatter plot of the ratio of Doppler parameter $b$ to the thermal Doppler
parameter $b_{\rm th}$ for the temperature of the gas cloud, as a function
of \HI column density. A substantial contribution to the Doppler parameter
is nonthermal, arising from bulk motions within the clouds. The cutoff along
the left of the distribution is due to selecting only systems with line center
opacity exceeding 0.05.}
\label{fig:bbth_HI}
\end{figure}
In Figure \ref{fig:specHI_od}, we show the relation between the absorption
features and the density contrasts responsible for them. Two effects are
particularly noteworthy. The overdensities are plotted on a velocity scale
attached to the comoving frame, while the spectra include the effects of the
peculiar velocity field of the baryons on the spectral features. Although
each spectral absorption feature may be identified with an upward baryon
fluctuation relative to its local background, the features do not always
line up in velocity space. The density enhancements that give rise to the
\Lya absorbers have systemic peculiar velocities of as much as a few hundred
kilometers per second.

The second effect is that many discrete absorption features are associated
with fluctuations that are {\it underdense} relative to the cosmic
average density, both in the baryons and the dark matter, as shown in
Figure \ref{fig:corr_NHI}. These features tend to be those that are optically
thin at line center, for which $\NHI<10^{13}\cm2$. Because the dark matter
density associated with the features tends to be below the cosmic average,
these structures appear not to be gravitationally bound. Yet, as shown in
Figures \ref{fig:corr_NHI} and \ref{fig:specHI_od}, no pronounced separation
between the baryons and the dark matter is found. In the following section,
we investigate these apparently anomalous structures found for the lowest
column density systems.

In Figure \ref{fig:bbth_HI} we show a scatter plot of the ratio of the
Doppler parameter of the \HI features to the value expected due to
thermal broadening alone, $b_{\rm th}=(2kT/m_{\rm H})^{1/2}$.  A
substantial contribution to the line--broadening derives from
nonthermal motion within the clouds. This may be due to internal
structure within the clouds that is unresolved in the \HI absorption
feature due to the finite velocity width of the lines. Hu \etal (1995)
argue for subcomponents based on the spread in $b$--parameters. Cowie \etal
(1995) similarly argue for subcomponents
based on the number of \ion{C}{4} features found in systems with
$\log\NHI>14.5$ (see also Songaila \& Cowie 1996). The simulations
typically find subcomponents in \ion{C}{4} as well, assuming a uniform
enrichment of carbon (Haehnelt \etal 1996; Zhang \etal 1997).

\subsection{\HI Opacity Distribution}

\begin{figure}
\plottwo{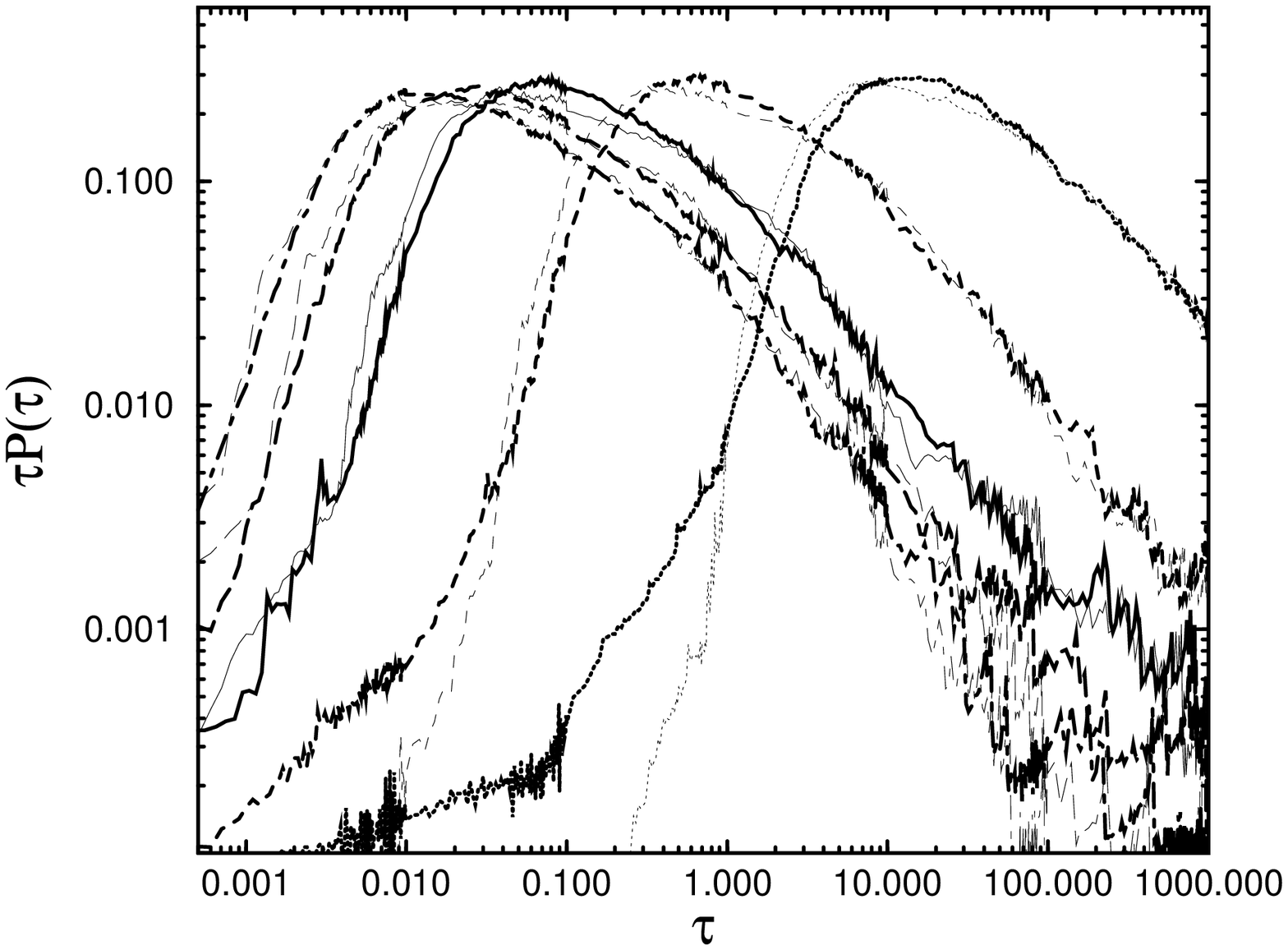}{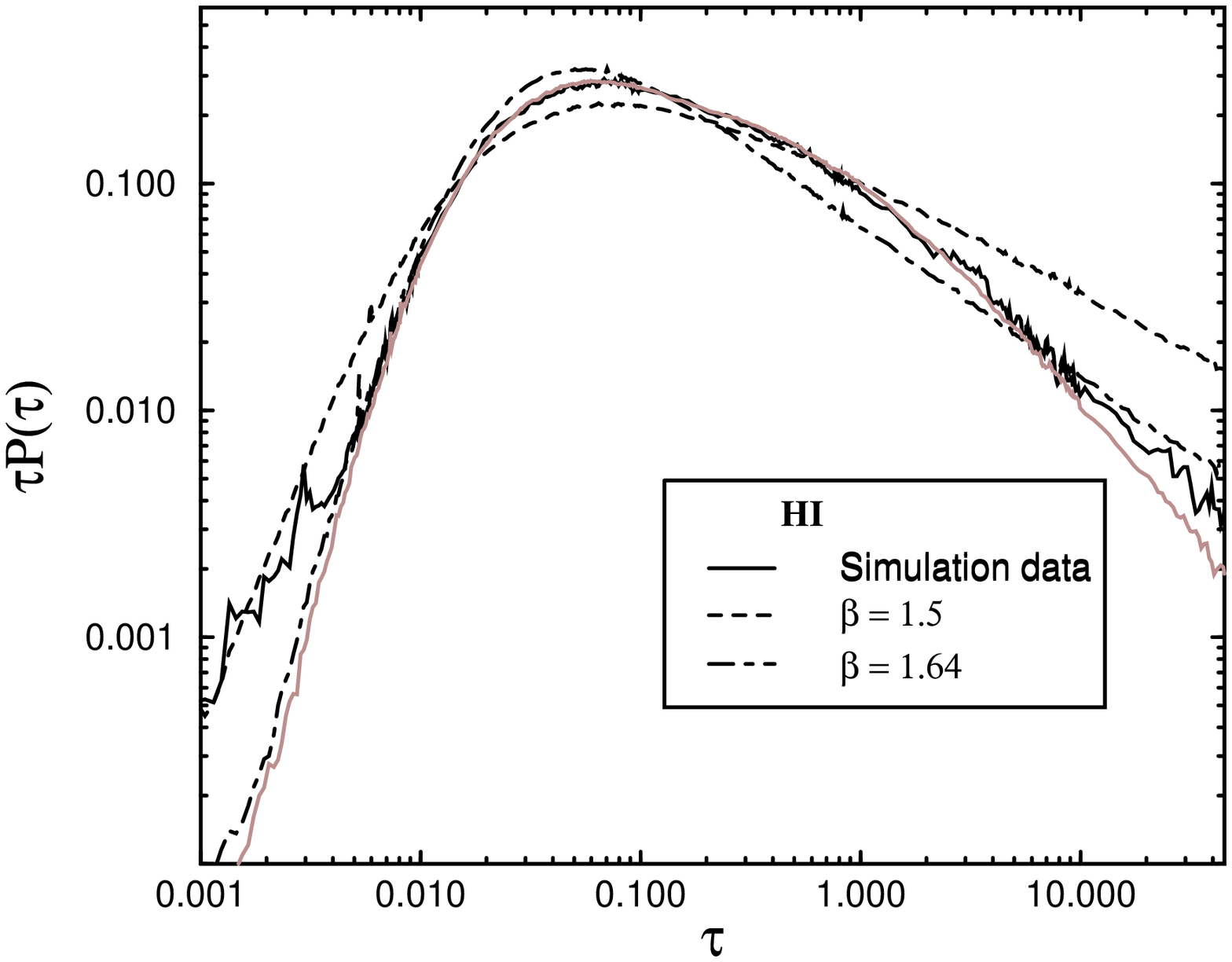}
\caption{ 
(a)\ The distribution of \HI opacity in the spectra at $z=5$ ({\it dotted}),
$z=4$ ({\it short--dashed}), $z=3$ ({\it solid}), $z=2.4$ ({\it long--dashed}),
and $z=2$ ({\it dot--dashed}). The heavy lines are for the top grid and
the light lines for the subgrid. The agreement between the top and subgrids
shows that the simulation has converged. The
subgrid opacity is somewhat lower because the subgrid was placed on an
underdense region.\ (b)\ The opacity distribution at $z=3$ from the simulation
({\it solid}), and for two line models with power law line center opacity
distributions $dN/d\tau_0\propto\tau_0^{-\beta}$, with the indicated values of
$\beta$. The thick gray line shows the distribution from a line model using the
measured $\tau_0$ distribution for $\tau_0>0.1$, and smoothly extended to
lower values using a power law with $\beta=1.55$. The opacity distribution
is well reproduced by a line model. No significant continuous Gunn--Peterson
component is allowed.}
\label{fig:ftau_HI}
\end{figure}

In Zhang \etal (1997), we showed that it was possible to account for
the mean intergalactic \ion{H}{1} and \ion{He}{2} \Lya absorption
entirely by discrete absorption systems identified in the simulated
spectra. No evidence for a significant residual opacity due to the
Gunn-Peterson effect was found. We now ask if the same is true of the
full opacity distribution of the spectra. In Figure \ref{fig:ftau_HI},
we show the probability distribution $P(\tau)$ for an optical depth
in the range $(\tau,\tau+d\tau)$ in a given pixel in the synthetic
spectra. The pixel widths are set at $0.6\kms$. In the left panel, we
show the evolution of the opacity distribution for both the top grid
and subgrid. Both because the subgrid is centered on an underdense
region and because it is a factor 64 smaller in volume, we do not
expect perfect correspondence with the top grid, but the agreement
between the top and subgrid results shows that we are resolving the
essential structures responsible for the spectral opacity.

In the right panel, we attempt to reproduce the distribution at $z=3$
from a discrete line model using the results found in Zhang \etal
(1997) for the distribution in line center opacity $\tau_0$ and
Doppler parameter $b$ of the \Lya forest lines identified in our
synthetic spectra, as given by Voigt profile fitting. There we found
that for $\tau_0>0.05$, the best power law fit to the line center
opacity distribution, $dN/d\tau_0\propto\tau_0^{-\beta}$, was given by
$\beta=1.64$. We adopt a lognormal distribution for the Doppler
parameters according to Zhang \etal of
$f(b)\propto\exp[-6.8\log^2(b/26.1)]$.  (We find that our results are
not substantially altered by choosing the best fitting gaussian
distribution instead.) Because the number of lines diverges at the low
end for a perfect power law, it is necessary to impose a lower cutoff
to the distribution. The resulting spectral opacity distribution
$P(\tau)$ for small values ($\tau\ll0.1$), is affected by the choice
of a minimum $\tau_0$. Decreasing the minimum forces the distribution
$\tau P(\tau)$ to turn down at increasingly larger values of $\tau$.
The reason for this behavior is that as the number of clouds is
increased, the occurrence of a low opacity excursion in a given pixel
becomes increasingly improbable:\ the forest tends to blanket the
spectrum everywhere. The extension of the distribution in Figure
\ref{fig:ftau_HI} to very low values (we still find a signal for
$\tau<0.001$), shows that the limit of a continuum in spectral
coverage is actually not reached. The distribution in $\tau_0$ does in
fact cut off.

Using the above distributions, we perform Monte Carlo realizations of the
spectra in order to determine if the spectral opacity distribution may be
accounted for entirely by a line model. We adjust the lower
cutoff for two power-law models, $\beta=1.64$ and $\beta=1.5$, in
order to obtain a good match to $P(\tau)$. We find each reproduces
different regions in $\tau$ well, but neither is completely satisfactory. We
next perform realizations drawing $\tau_0$ from the actual line center opacity
distribution we obtain from the simulated spectra. Since the distribution
becomes incomplete for low values, we use the measured distribution only for
$\tau_0>0.1$. We then extrapolate the distribution to lower values using a
power law. We are now able to reproduce the spectral opacity distribution
$P(\tau)$ to high precision, as shown by the thick gray curve in
Figure \ref{fig:ftau_HI}. The fit shown is for $\beta=1.55$ and
$\tau_0>5\times10^{-4}$ for the low $\tau_0$ extension. For the adopted mean
Doppler parameter, the lower cutoff corresponds to a neutral \HI column density
of $\NHI=1.7\times10^{10}\cm2$. Because a sharp cutoff is artificial, this
value should only be taken as an indication of the magnitude in column density
below which relatively few systems form. The deviations at
the extremities in the distribution we believe are due to sampling error
and possible curvature in the $\tau_0$ distribution at low values. We thus
conclude that the full spectral opacity distribution $P(\tau)$ may be
accounted for entirely by line blanketing due to discrete absorption systems
with Voigt line profiles. If a smoothly distributed, homogeneously expanding
\ion{H}{1} component were present, the spectral opacity distribution would
cut off abruptly at the opacity corresponding to its \ion{H}{1} density. Figure
\ref{fig:ftau_HI} shows no such component exists.

We may compare the distribution at $z=2.4$ with that of Croft \etal (1997)
at $z=2.33$ (their Figure 11), based on a TreeSPH calculation.
While the two distributions agree in shape for $\tau>0.05$, we find a tail at
lower values that is absent in the Croft \etal distribution, which cuts off
at $\tau<0.004$. The tail is a signature of the low opacity regions in the
spectrum that result from the ubiquitous clumping of the gas. A comparison
between Figure \ref{fig:specHI_od} and Figure 4 of Croft \etal shows that
we obtain substantially more structure in the transmitted flux and underlying
gas distribution than in the TreeSPH simulation. In \S\ref{subsubsec:size}
above, we showed that the characteristic cloud sizes and masses were
100 kpc and $M_c\approx10^9\msun$. The mass of the gas particles in
the simulation of Croft \etal is $1.5\times10^8\msun$, not much less
than the cloud masses. We believe that the reason the low opacity structure
is absent in their simulation is that they are under-resolving the clumping
of the gas. Under--resolving the clumping will contribute to the higher
opacity values they obtain compared to our results (Zhang \etal 1997).

\section{Cosmic Minivoids}
\label{sec:minivoids}

In this section, we discuss in greater detail the absorption and
evolution of discrete systems in the underdense regions. We believe it
illuminating to distinguish these systems from the remainder of the
\Lya forest for several reasons:\ 1.\ We find considerable structure
in the underdense regions, more than can be accounted for by the Jeans
instability. 2.\ The systems are diffuse, and hence behave similarly
to a smooth IGM component in their averaged absorption properties, but not
identically to one. 3.\ The recent and anticipated space--based
detections of intergalactic \HeII essentially are probes of the fine
structure of the minivoids. Since the current and near--term detectors
are not generally capable of resolving individual \HeII \Lya absorption
systems, a discussion of
the underlying physical properties of the clouds in the underdense
regions is crucial for interpreting future \HeII measurements in the
context of CDM and reionization models. Critical to the discussion is
establishing convergence to the correct amount of \HeII absorption.
Without adequate resolution of the structures giving rise to the absorption,
the amount of absorption will be over--estimated.

\subsection{Physical Structure}
\label{subsec:mv_struc}

The low physical densities associated with the low column density \HI
systems was previously demonstrated in Figures \ref{fig:cont_z3} and
\ref{fig:cont_col}. In Figure
\ref{fig:specHI_od}, we show a direct comparison between the \HI
absorption and the associated underlying baryon and dark matter
density fluctuations at $z=3$. For \HI column densities below about
$10^{13}\cm2$, lines that are optically thin in \Lya at line center,
the typical baryonic and dark matter densities of the systems are below
the cosmic average. The low densities present a difficulty for accounting
for their origin, since such low density systems appear not to have
formed from a Jeans instability. The baryonic
proper Jeans length in a medium of mixed dark matter and baryons is
$\lambda_J=2\pi(2/3)^{1/2}c_s/ H(z)$, where $c_s$ is the sound
speed of the baryons associated with a linear perturbation and $H(z)$
is the Hubble constant at redshift $z$ (e.g., Bond \& Szalay 1983).
For an isothermal perturbation, $\lambda_J\simeq1\,{\rm
Mpc}\,h_{50}^{-1} T_4^{1/2}(1+z)^{-3/2}$, or about 150~kpc at $z=3$.
This gives a minimal column density due to Jeans instability of $\log
N_{\rm HI} \simeq13$ at $z=3$. The corresponding baryonic Jeans mass
is $M_J\equiv\langle\rho_b\rangle\lambda_J^3\simeq4\times10^{9}\msun
(\Omega_b h_{50}^2/0.06)h_{50}^{-3}T_4^{3/2}
(1+z)^{-3/2}$, or $\sim4\times10^8\msun$ at $z=3$.
The power--spectrum of the density fluctuations for short wavelength
modes, $\lambda\ll\lambda_J$, will be suppressed by the factor
$(\lambda/\lambda_J)^2$. Thus, one might reasonably expect a downturn
in the \HI column density distribution below $\sim10^{13}\cm2$.

\begin{figure}
\plottwo{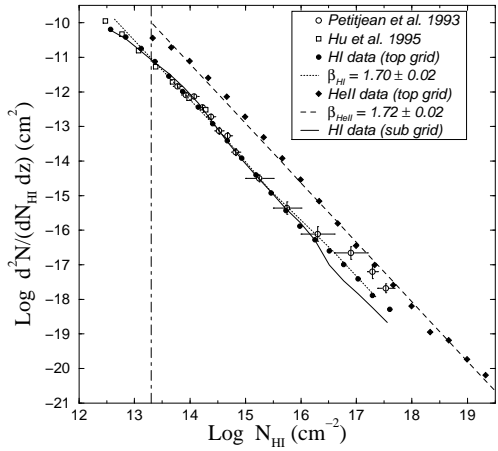}{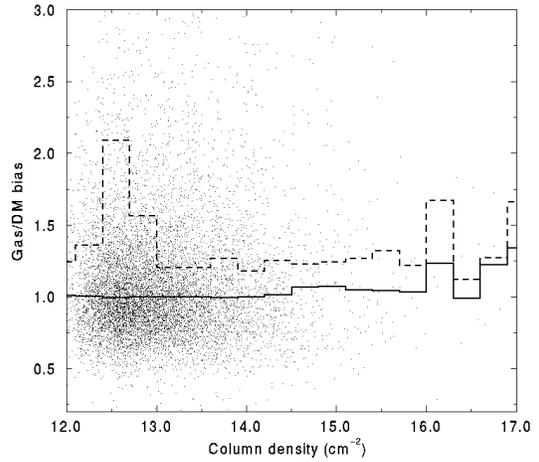}
\caption{
\HI (filled circles) and \HeII (filled diamonds)
column density distributions at $z=3$. Also shown are the observed data
from Petitjean \etal (1993) and Hu \etal (1995). The dotted and dashed lines
are the least--squares power law fits to the entire
column density ranges of \HI and \HeII respectively.
The solid line represents the \HI distribution derived from data on
the higher resolution subgrid. The subgrid is centered on an underdense
region in the box, and so it misses the highest column density systems.
The vertical dashed line indicates
$\NHI=2\times10^{13}\cm2$, the dividing line between the
optically thin and thick components.}
\label{fig:dNdNHI}
\caption{
Scatter plot of the ratio of gas to dark matter overdensities as
a function of \HI column density, at $z=3$.}
\label{fig:obDM_NHI}
\end{figure}
This line of argument was pursued by Reisenegger \& Miralda-Escud\'e
(1995), who concluded that the diffuse optically thin systems
would merge into a `fluctuating Gunn-Peterson effect.' No indication of a
suppression in the column density distribution at $10^{13}\cm2$
is apparent in Figure \ref{fig:dNdNHI} (reproduced from Zhang \etal 1997):\ the
power--law column density distribution persists to column densities
well into the optically thin regime, with a slope consistent with
measurements of the \Lya forest by the Keck HIRES (Hu \etal 1995).
Figure \ref{fig:corr_NHI} shows that the low column density absorbers are
associated with structures that are underdense both in the baryons and
the dark matter distributions. If they formed from Jeans unstable fluctuations,
the opposite would be true. Thus {\it we find discrete absorption systems that
appear not to have formed from a Jeans instability.}
It is crucial to resolve the low column density systems for assessing the
amount of \HeII absorption predicted by the model. We would therefore like
to understand the origin of these systems to ensure that we are able to
converge to the correct average intergalactic \HeII opacity.

A clue to the formation mechanism of the highly diffuse clouds is
suggested by a comparison of the low \HI column density contours in
Figure \ref{fig:cont_col} with the peculiar velocity divergence
contours in Figure \ref{fig:cont_z3}. The low \HI column density
systems tend to associate with regions of positive peculiar velocity
divergence. The optically thin systems are not equilibrium
structures:\ their density is dissipating at a rate in excess of the
average cosmic expansion. In principle, low column density gas could
be associated with low mass minihalos. Bond \etal (1988) demonstrated
that the baryons bound to low mass minihalos prior to reionization may
be reheated to temperatures too high for the minihalos to retain the
gas after reionization, and the baryons will escape. While this must
occur for some systems, a comparison with the dark matter overdensity
for the optically thin features in Figures \ref{fig:corr_NHI} and
\ref{fig:specHI_od} show that these systems are generally not associated
with minihalos, since the dark matter fluctuations themselves are below
the cosmic average. Statistically, no substantial separation of the baryons
from the dark matter is found for these low column density systems,
as shown in Figure \ref{fig:obDM_NHI}.

\begin{figure}
\plottwo{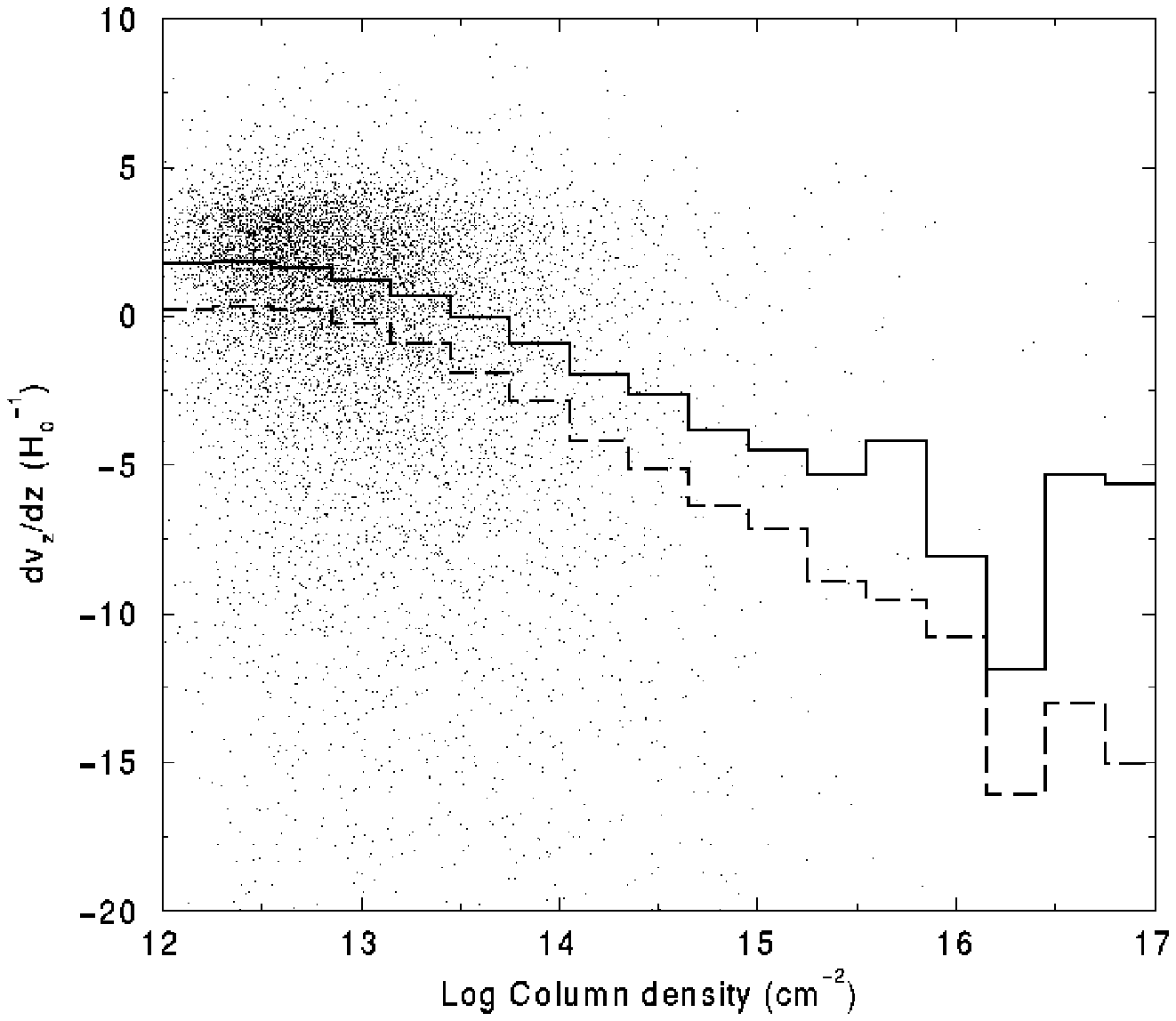}{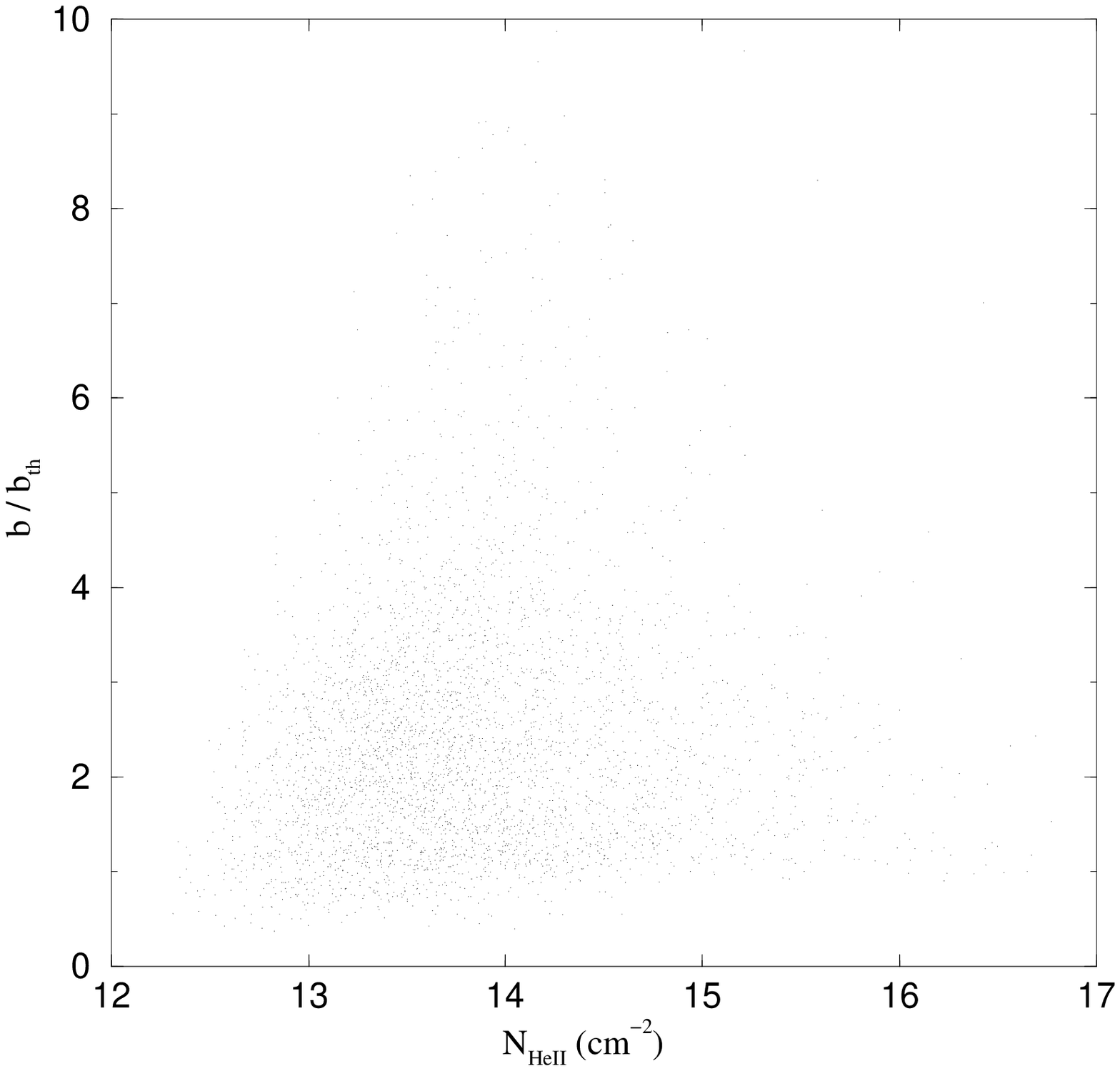}
\caption{
Scatter plot of the line--of--sight peculiar velocity divergence as
a function of \HI column density, at $z=3$.}
\label{fig:dvzdz_NHI}
\caption{
A scatter plot of the ratio of \HeII Doppler parameter $b$ to the thermal
Doppler parameter $b_{\rm th}$ for the temperature of the gas cloud, as a
function of \HeII column density. A substantial contribution to the Doppler
parameter is nonthermal, arising from bulk motions within the clouds. The
cutoff along the left of the distribution is due to selecting only systems
with line center opacity exceeding 0.05.}
\label{fig:bbth_HeII}
\end{figure}
Another possibility is that the features arise from velocity caustics,
regions for which a convergence in the line--of--sight velocity field
compresses the absorption in redshift, hence wavelength, space (McGill 1990).
This is possible even in low density regions where the flow field is
divergent, since the gas may be expanding in the directions lateral to
the line--of--sight. In Figure \ref{fig:dvzdz_NHI} we show a scatter plot
of the line--of--sight velocity derivative $dv_z/dz$ as a function of $\NHI$,
where $v_z$ is the component of the peculiar velocity along the
line--of--sight. The low column density absorbers tend to have $dv_z/dz>0$,
opposite to the criterion required for velocity caustics.

Meiksin (1997) has suggested an interpretation of these features in
terms of the growth of fluctuations in an underdense background. A fluctuation
that is underdense compared to the mean cosmological value but overdense
relative to a large surrounding underdense region will grow as if in an
open universe. After an initial period of growth relative to the
diminishing local background, the relative density perturbation will
`freeze', very roughly at an epoch given by
$1+z_f\approx\Omega_v^{-1}-1$, where $\Omega_v$ is the ratio of
density in the background void to the Einstein--deSitter critical
density. Thus, although the physical density of the absorber
diminishes with time like the density of the background void, it
retains its integrity as a discrete entity as the void continues to expand.

\subsection{Optically Thin \HI Absorption}
\label{subsec:mv_HI}

The resonant opacity arising from a uniform medium of \ion{H}{1} density
$\bar n_{\rm HI}$ in a homogeneously expanding universe is given by
(Field 1959; Gunn \& Peterson 1965)
\begin{equation}
\tau_\alpha=\left(\frac{\pi e^2}{m_e c}\right)f_\alpha\lambda_\alpha
\frac{1}{H_0}\frac{\bar n_{\rm HI}}{(1+z)(1+2q_0z)^{1/2}},
\label{eq:tau_thin}
\end{equation}
where $f_\alpha$ is the upward oscillator strength for \Lya, and
$\lambda_\alpha$ is the \Lya rest wavelength. (A zero cosmological
constant is assumed, so that $q_0 =\Omega_0/2$.) What is the effect on
the opacity of aggregating the gas into discrete clouds? Provided that the
only clouds considered are those that are individually optically thin, the
expression for the opacity is unchanged if $\bar n_{\rm HI}$ is identified
with the spatially averaged density of the neutral hydrogen.
This is distinct from the average internal neutral density $n_{\rm
HI}^c$ of the individual clouds when their volume filling factor $f_c$
is less than unity (e.g., Meiksin 1997). The two are related by $\bar
n_{\rm HI}=f_c \bar n_{\rm HI}^c$. Eq. (\ref{eq:tau_thin}) shows that
the optically thin component of the \Lya forest behaves similarly, but
not identically, to a diffuse homogeneous gas component. While the
opacity in both cases is in direct proportion to the spatially
averaged neutral hydrogen density, and so inversely proportional to
the metagalactic UV radiation field, the opacities need not evolve in
the same way, both because the cloud internal density need not
evolve like the cosmological expansion, and because of any
evolution of the filling factor.

The presence of the volume filling factor precludes inverting
eq. (\ref{eq:tau_thin}) to solve for the spatially averaged total
hydrogen density of the optically thin systems. If
the internal density of the clouds, however, is less than the cosmic
mean density, then we may derive a lower limit to the cosmic mean
density given a measurement of $\tau_\alpha$ from the optically thin
systems. (Since the result that the optically thin clouds are
underdense is obtained from the simulations, this is actually a
consistency check on the simulation results rather than an independent
determination of $\Omega_b$.) The sample of Hu \etal (1995) may be
used for making such a determination. Counting only lines with line
center \Lya opacity less than 0.5, to be conservative, and weighting
each line inversely by the estimated incompleteness factor for its
\ion{H}{1} column density according to their Table 3, we obtain
$\tau_\alpha\approx0.07$. Combining with eq. (\ref{eq:nHI}), and
requiring $\rho^c_b < \bar \rho_b$, where $\rho^c_b$ is the internal
cloud baryon density and $\bar \rho_b$ is the average cosmological
baryon density, we obtain $\Omega_b>0.02h_{50}^{-3/2}$, consistent
with the value $\Omega_b=0.06$ adopted in the simulation.

\subsection{Intergalactic \HeII Absorption}
\label{subsec:mv_HeII}

A consequence of the low nonequilibrium temperature of the gas
inside the minivoids is that the
temperature of the systems giving rise to the low column density
absorbers ($\NHI<10^{13}\cm2$), will be history--dependent; i.e., it
will depend on the photoionization history of the baryons and their
expansion history, which in turn depends on the local density
inhomogeneities. The temperature will depend on the actual density of
the gas as well. In this case, the Doppler parameters of the low column
density lines are no longer independent of $b_{\rm ion}$.
This is especially important for determining the amount of
\HeII absorption, since it sets the widths of the absorption features,
hence the total amount of absorption once the lines begin to enter the
saturated part of the curve--of--growth. In practice, the effect is
reduced by the presence of nonthermal broadening of the lines due to
bulk motions. In Figure \ref{fig:bbth_HeII}, we show a scatter plot
of the ratio of the Doppler parameters to the thermal Doppler parameter
at the cloud temperature, as a function of $N_{\rm HeII}$. A large
contribution to the widths of the lines derives from bulk motion within
the clouds.

\begin{figure}
\plottwo{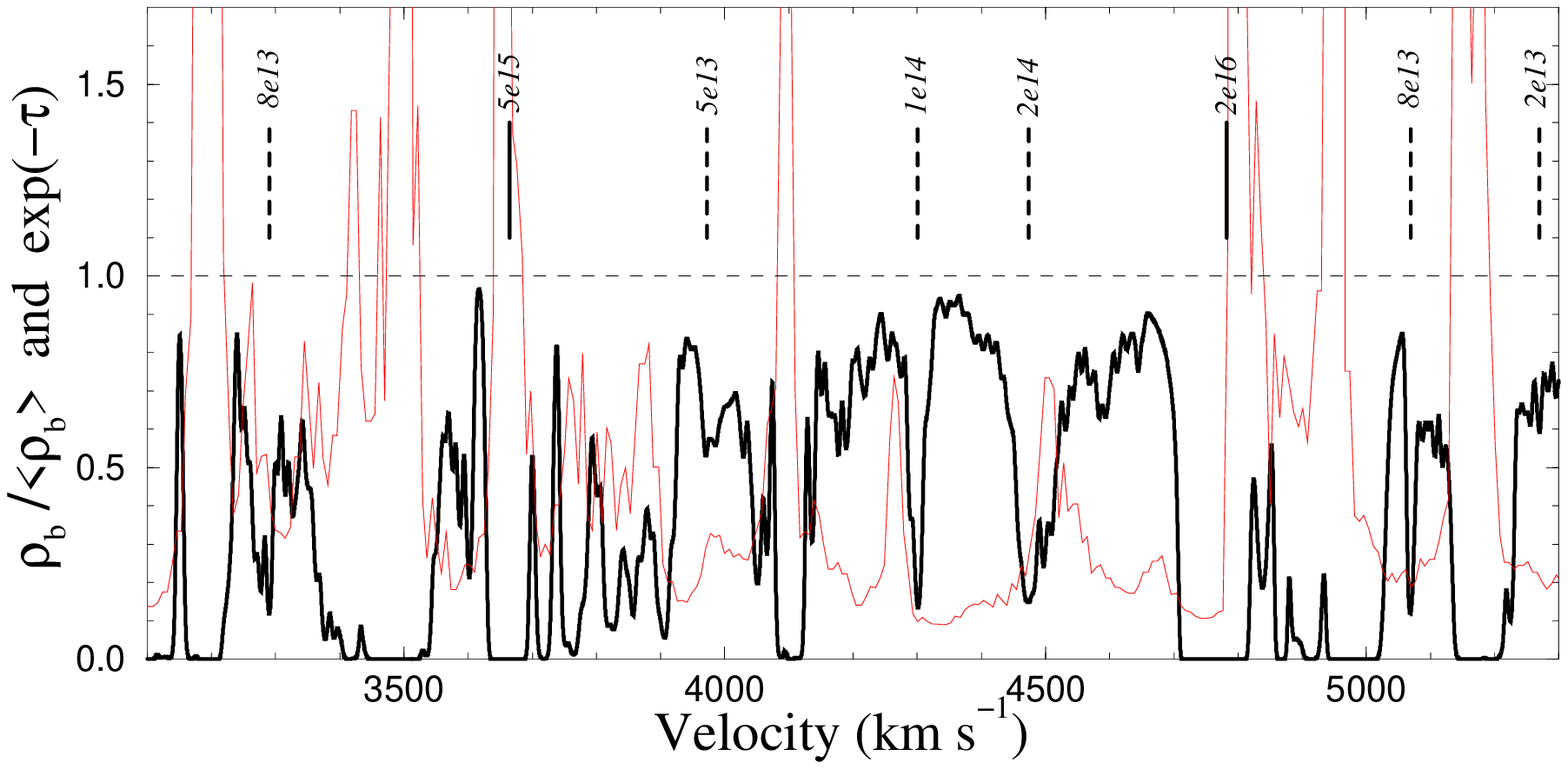}{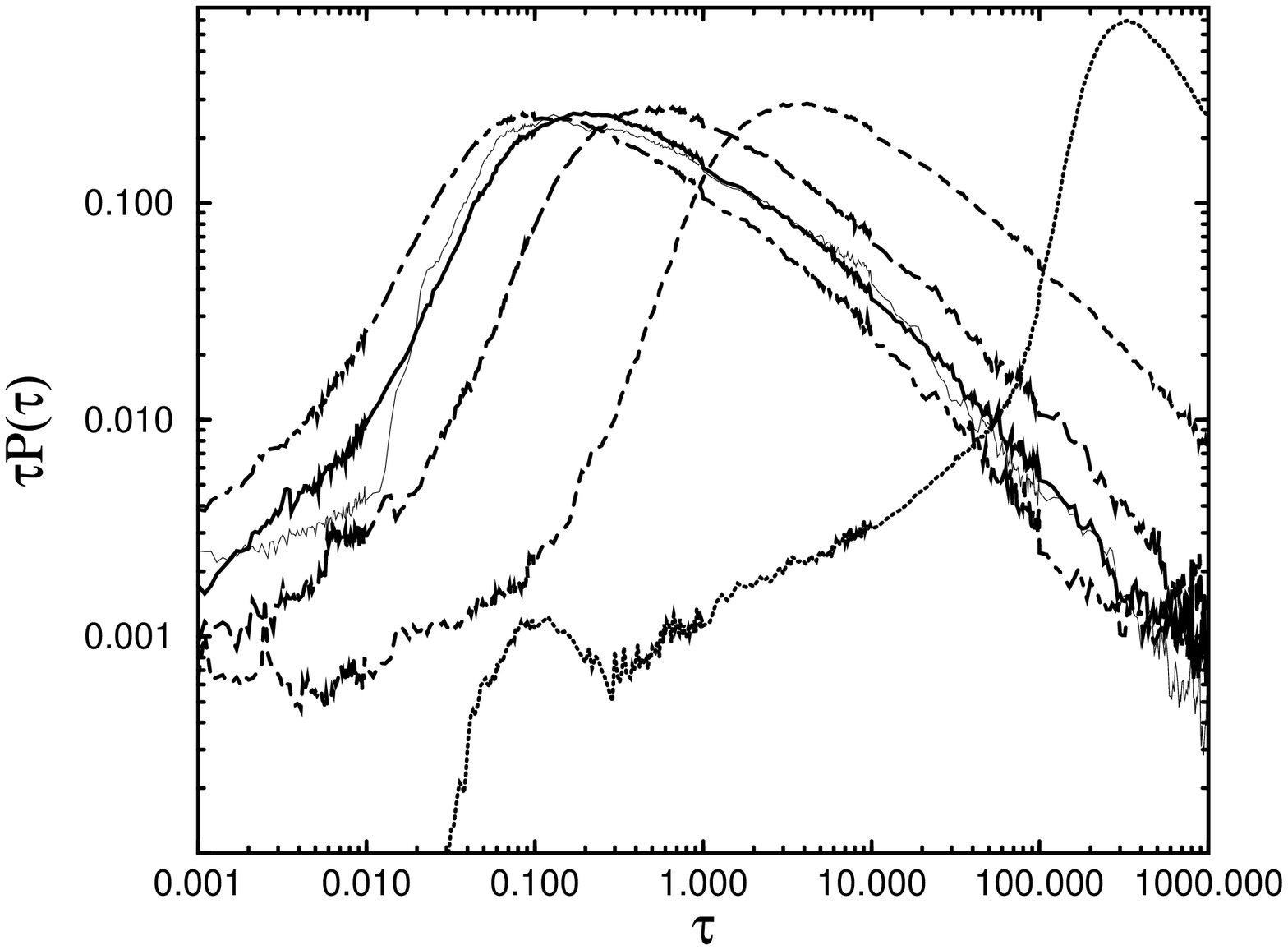}
\caption{
A sample of the \HeII absorption spectrum along a line--of--sight through the
box at $z=3$. Also shown ({\it light line}) is the baryonic overdensity that
gives rise to the absorption features. Most of the discrete
\HeII systems arise from density fluctuations in underdense regions.}
\label{fig:specHeII_od}
\caption{
The distribution of \HeII opacity in the spectra at $z=5$ ({\it dotted}),
$z=4$ ({\it short--dashed}), $z=3$ ({\it long--dashed}), $z=2.4$, both for
the top grid ({\it heavy solid}), and the subgrid ({\it light solid}), and
$z=2$ ({\it dot--dashed}). The agreement between the top and subgrid
distributions shows that the simulation is resolving the essential features
that determine the mean intergalactic \HeII opacity.}
\label{fig:ftau_HeII}
\end{figure}
In Figure \ref{fig:specHeII_od} we show a representative \HeII
spectrum at $z=3$. As for the \HI, the spectral features tend to be
associated with discrete baryonic density fluctuations. Because of the
higher density of \HeII compared to \HI, however, these features are
largely associated with fluctuations in the underdense regions:\ \HeII
absorption provides a probe of the cosmic minivoids. Because of the
sensitivity of the void temperature and baryonic density structure to
the rate of expansion within the voids and the ionization history, the
\HeII opacity may provide a useful means of discriminating between
rival cosmological models and reionization scenarios. To do so,
however, it is crucial that the structure within the voids be resolved
sufficiently to ensure convergence to the correct total \HeII opacity.

We show in Figure \ref{fig:ftau_HeII} the evolution of the
distribution function $P(\tau)$ of the \HeII spectral opacity from the
top grid calculation for $2<z<5$. At $z=2.4$, we also show the opacity
distribution from the subgrid calculation. The subgrid and top grid
distributions agree well, showing that we have resolved the features
that produce the \ion{He}{2} opacity. The opacity is somewhat lower in
the subgrid because we have centered it on a low density region. We
found in Zhang \etal (1997) that matching to the intergalactic
\ion{He}{2} opacity measurements required a break in the UV background
intensity between the \ion{H}{1} and \ion{He}{2} photoelectric edges by
a factor of 100--150. This is consistent with the break required to match the
amount of Si and C absorption measured in the \Lya forest by Songaila
\& Cowie (1996) (\S \ref{subsec:cloudionization}).

\section{Summary}
\label{sec:summary}
We summarize our main results.

\noindent 1.\ The structures giving rise to the \Lya forest cloud
population are characterized by a range of morphologies, with
different structures associated with different ranges in \ion{H}{1}
column density. The high column density absorbers ($\NHI>10^{15}\cm2$
and $\NHeII>10^{17}\cm2$) correspond to the highly overdense
structures ($\rho_b/\overline{\rho}_b >10$) residing mostly along and
at the intersections of filaments. The medium column density absorbers
($10^{13}\lsim\NHI\lsim10^{14}\cm2$ and $10^{15}\lsim\NHeII\lsim10^{16}\cm2$)
correspond to the modestly overdense filaments
($1<\rho_b/\overline{\rho}_b<5$). The column density can be coherent
over the scale of a few megaparsecs at this level. The lowest column
density absorbers ($\NHI\sim10^{12}\cm2$ and $\NHeII\sim10^{14}\cm2$)
are associated with underdense structures
($\rho_b/\overline{\rho}_b<1$) and are located in the void regions
between the filamentary structures. They are typically a few hundred
kiloparsecs across. The associated flow pattern of the structures is
one of outflow from the voids, compression into sheets at the
boundaries of the voids, and a resulting flow along the sheets toward
their intersections, where the densest structures form.

\noindent 2.\ The principal density structures are in place by $z=5$,
with only moderate evolution in the baryon overdensity at later times.
By contrast, the post--reionization temperature is initially narrowly peaked
at the photoionization value, and flattens toward higher and lower
values with time as dense massive structures continue to collapse and
the voids continue to cool by expansion.

\noindent 3.\ The baryon overdensity, dark matter overdensity, baryon
temperature, and divergence of the baryon peculiar velocity field all show
strong correlations with the \ion{H}{1} column density of the associated
absorption features. For $12.5<\log\NHI<14.5$ at $z=3$, we obtain
$\rho_b/\langle\rho_b\rangle\approx N_{\rm HI,13}^{1/2}$,
$T_4\approx0.8 N_{\rm HI,13}^{1/4}$, and $f_{\rm HI}\approx5\times10^{-6}
N_{\rm HI,13}^{5/16}$, where $N_{\rm HI,13}$ is the \ion{H}{1}
column density in units of $10^{13}\cm2$, and $T_4$ is the baryon
temperature in units of $10^4$~K. The baryonic density scaling requires
that all clouds in this column density range have the same characteristic
size of $\NHI/n_{\rm HI}=100-150$ kpc. The associated cloud mass is
$0.3-3\times10^9\msun$.

\noindent 4.\ The simulation is able to reproduce the statistics of the carbon
and silicon measurements of the \Lya forest at $z\sim3$ of Songaila \& Cowie
(1996) and the intergalactic \ion{He}{2} opacity measurements, provided the UV
background has a break of $\Gamma_{\rm HI}/\Gamma_{\rm HeII}\approx250-400$
and the Si to C abundance ratio is a few times the solar value. Combining with
the limits on $b_{\rm ion}$ from our spectral analysis (Zhang \etal 1997), we
obtain for best estimates of the cosmic mean baryon density and UV background
$0.03\lsim\Omega_b\lsim0.08$ ($h_{50}=1$), and
$0.3\lsim\Gamma_{\rm HI,-12}\lsim1$ at $z\approx3-3.5$. The constraints on the
radiation field are consistent with a UV background dominated by QSO sources
with a spectral index of $\alpha_Q\approx1.8-2$, in agreement with the indices
measured by Zheng \etal (1996). These values are sensitive to the assumed
normalization and shape of the primordial power spectrum.

\noindent 5.\ We find that half of the baryons in the simulation are contained
within clouds with \ion{H}{1} column densities in the range
$12.5<\log\NHI<14$ at $z=3$, and that fewer than 5\% reside in systems that
have not been identified with discrete absorption lines in the synthesized spectra.

\noindent 6.\ The structures giving rise to the absorption systems have
systemic peculiar velocities as high as $100\kms$ and internal motions that
give substantial contributions to the Doppler widths of the absorption lines.

\noindent 7.\ The \ion{H}{1} opacity distribution of the synthesized spectra
may be fully accounted for by a distribution of discrete absorption lines. No
significant uniform Gunn--Peterson component is allowed.

\noindent 8.\ A large population of optically thin absorption lines is
associated with underdense modulations in minivoids, low density regions
a few megaparsecs across. Most of the intergalactic \ion{He}{2} opacity
is derived from absorption within the minivoids. The dark matter fluctuations
associated with this absorber population are underdense, and the peculiar
velocity field of the baryons is divergent, suggesting that the systems did
not originate through a Jeans instability. They appear to have grown from small
scale primordial underdense fluctuations within the larger minivoid.
If the optically thin absorbers measured in QSO spectra are
associated with underdense regions, we derive a lower bound on the cosmic
baryon density of $\Omega_b h_{50}^{3/2}>0.02$.

\acknowledgements
We are pleased to thank Ed Bertschinger, Piero Madau, David Weinberg, and
Martin White for useful conversations, and Francesco Haardt for providing
us with QSO dominated UV background spectra.
We would also like to thank John Shalf (NCSA), for his 3D visualizations
in the accompanying video tape and in Plate 1.
This work is supported in part by the NSF under the auspices of the
Grand Challenge Cosmology Consortium (GC$^3$). The computations
were performed on the Convex C3880 and the SGI Power Challenge at the
National Center for Supercomputing Applications, and the Cray C90 at
the Pittsburgh Supercomputing Center under grant AST950004P. A.M. thanks
the William Gaertner Fund at the University of Chicago for support.



\clearpage

\noindent Plate 1 -- Isodensity contour surfaces of log baryon overdensity
($\rho_b/\bar\rho_b$), at $z=3$. The contour levels are
$\log_{10}(\rho_b/\bar\rho_b)=-1$, --0.5, --0.3, 0., 0.5, and 1.
The colors indicate temperature,
ranging from $10^4$~K ({\it blue}) to $10^5$~K ({\it red}).

\end{document}